\documentclass[pra,superscriptaddress,amsmath,amssymb,twocolumn]{revtex4-2}
\usepackage{graphicx}
\usepackage{subfigure}
\usepackage{adjustbox}
\usepackage{bm}
\usepackage{bbm}
\usepackage{color}
\usepackage{braket}
\usepackage{standalone}
\usepackage{multirow}
\usepackage{tikz}
\usepackage{mathrsfs}
\usepackage{dsfont}
\usepackage[colorlinks,bookmarks=true,citecolor=blue,linkcolor=blue,urlcolor=blue]{hyperref}
\usepackage{cleveref}
\usepackage{comment}
\usepackage{mathtools}
\usepackage{soul}

 \Crefname{equation}{Eq.}{Eqs.}
\Crefname{figure}{Fig.}{Figs.}

\begin{document}
\title{Interaction quench of dipolar bosons in a one-dimensional optical lattice}

\author{Paolo Molignini}
\affiliation{Department of Physics, Stockholm University, AlbaNova University Center, 10691 Stockholm, Sweden}
\author{Barnali Chakrabarti}
\affiliation{Department of Physics, Presidency University, 86/1 College Street, Kolkata 700073, India}
\date{\today}

\begin{abstract}
A Tonks-Girardeau (TG) gas is a highly correlated quantum state of strongly interacting bosons confined to one dimension, where repulsive interactions make the particles behave like impenetrable fermions. 
By suddenly tuning these interactions to the attractive regime, it is possible to realize a super-Tonks-Girardeau (sTG) gas -- a highly excited, metastable state of strongly attractive bosons with unique stability properties.
Inspired by the sTG quench scenario, we investigate a similar setup but with the inclusion of long-range dipolar interactions, which modify the system away from the TG Mott insulating limit. 
We simulate an interaction quench on dipolar bosons initially prepared in various states and fillings, using real-space densities, orbital occupations, Glauber correlation functions, and autocorrelation functions to probe post-quench stability. 
Our results reveal that stability is maintained only at very weak dipolar interaction strengths when starting from a unit-filled TG Mott state. 
In contrast, all cluster states -- whether unit-filled or doubly-filled -- eventually collapse under attractive interactions. 
This collapse is not always visible in the density profile but becomes apparent in the autocorrelation function, indicating complex many-body restructuring of the quantum state.
Our findings underscore the potential of dipolar interactions to drive novel quantum dynamics and highlight the delicate balance required to stabilize excited states in long-range interacting systems.
\end{abstract}
\maketitle

%%%%%%%%%%%%%%%%%%
%%%%% INTRO %%%%%
%%%%%%%%%%%%%%%%%%
%%%%%%%%%%%%%%%%%%%%%%%%%%%%%%%%%%%%%%%%%%%%%%%%%%%%%%%%%%%%%%%%%%%%%%%%%%%%%%%%%%%%%%%%%%%%%
\section{Introduction} 
Rapid experimental advances in controlling ultracold atoms and their interactions are making it possible to realize and study enticing states of matter that encapsulate the exotic properties of quantum mechanics.
These range from equilibrium phases such as supersolids and dipolar supersolids~\cite{Leonard:2017, Leonard:2017-2, Li:2017, Boettcher:2019, Tanzi:2019, Tanzi:2019-2, Chomaz:2019, Guo:2019, Natale:2019, Tanzi:2021, Norcia:2021, Sohmen:2021,  Sanchez-Baena:2023, Recati:2023}, to dynamical phenomena such as time crystals~\cite{Smits:2018, Kessler:2021, Kongkhambut:2022} and various Floquet phases~\cite{Aidelsburger:2015, Schweizer:2019, Li:2019, Wintersperger:2020, Bracamontes:2022, Sun:2023, Zhang:2023}.
Among these numerous recent advances, the super-Tonks-Girardeau (sTG) gas deserves special attention.
The sTG is a quantum state of strongly attractive bosons confined to a one-dimensional (1D) geometry~\cite{Astrakharchik:2005, Batchelor:2005, Tempfli:2008, Haller:2009, Kao:2021}, and it is remarkable because it is metastable despite being a highly excited state of matter~\cite{Boronat:2008,Astrakharchik:2008}. 
The sTG gas can be obtained from the standard Tonks-Girardeau (TG) gas~\cite{Tonks:1936, Girardeau:1960, Girardeau:1965, Olshanii:1998, Petrov:2000, Girardeau:2001, Kinoshita:2004, Paredes:2004, Yukalov:2005, Jacqmin:2011} under sudden switching of contact interactions from the strongly repulsive to the strongly attractive regime. 
The sTG state resembles a highly excited gas-like state and, in principle, it should have an infinite life time as it is an eigenstate of the Hamiltonian. 
Experimentally, it is in fact demonstrated that for short-range interactions the sTG state is a long-lived excitation.
Besides providing a novel paradigmatic behavior, the realization of the sTG phase opens up the possibility of creating other quantum states of ultracold quantum gases with similar counter-intuitive properties~\cite{Guan:2010, Murmann:2015, Barbiero:2015}.

In particular, in recent experiments with 1D arrays of bosons, the stability of the sTG phase has been revisited by adding a weak dipole-dipole interactions (DDIs) between the atoms. 
In the experiment of Kao {\it et. al.}~\cite{Kao:2021}, the robustness of the sTG phase against collapse has been studied for a bosonic one-dimensional quantum gas of dysprosium atoms. 
Unexpectedly, even though the DDIs break the integrability, they enhance the stability of the sTG phase in comparison to the short-range interaction case.
This observation raises the prospect of employing weak dipolar forces to influence stability without changing the energy of the sTG phase significantly.
This issue has been addressed recently by solving exactly a system of three trapped atoms with both contact and dipolar interactions~\cite{Chen:2023}, for which a distinct spectral response consisting of weaker hybridization with excited bound states is observed when repulsive DDIs are present.
Another recent work~\cite{Morera:2023} explores instead the unconventional mechanism of liquid formation in 1D optical lattice with strong on-site repulsion and weak long-range interactions. 
The underlying mechanism for liquid formation is established by a superexchange process. 
This discovery further prompted the investigation of the sTG quench in the corresponding geometry by means of extended Bose-Hubbard models, exploring the disruption of the states within a specific range of interactions~\cite{Marciniak:2023}. 
In this setup, strong local interactions are supplemented by weaker and attractive nonlocal interactions which lead to i) a gas phase, ii) a liquid phase and a iii) a self bound Mott phase. 
In the post quench dynamics, the liquid droplets expand and eventually evaporate. 

Our work aims to describe a similar interaction quench in an experimentally realistic continuum realization with bosons trapped in an optical lattice. 
We study richer initial setups, spanning more diverse states and particle fillings.
This allows us to unravel unexplored many-body mechanisms in the dynamics.  
We characterize the post quench metastable states with specific calculations of their life times and measures of correlation and many-body coherence.
We also elaborate on how the stability of the state can be controlled by the additional weak dipolar interactions.

Our initial conditions span clean Mott states, cluster states, and single-site localized quantum droplets with a) unit filling and b) double filling.
For each case, the short-range on-site repulsion is supplemented by long-range dipolar interactions which correspond to different initial states and dynamical scenarios.
We present many-body dynamics results by solving the time-dependent many-boson Schr\"odinger equation utilizing the MultiConfigurational Time-Dependent Hartree method for bosons (MCTDHB)~\cite{Streltsov:2006,Streltsov:2007,Alon:2007,Alon:2008,Lode:2016,Fasshauer:2016,Lode:2020}, implemented in the MCTDH-X software~\cite{Lin:2020,MCTDHX}. 
We extract key measures such as one- and two-body density dynamics, natural occupations, one- and two-body Glauber correlation functions, and autocorrelation functions.

Our overall observations are as follows: 
The TG gas without dipolar perturbation and with very weak dipolar attractions -- which correspond to scattering states -- remains stable for a long time under the interaction quench, as expected. 
However, in contrast to the findings in Bose-Hubbard models, all weakly bound Mott states and cluster states eventually deviate from the initial state and even collapse completely.
This is particularly true for fragmented quantum droplets where all the particles are concentrated in the same site.
The one-body and two-body correlations reveal the details of this destabilization, with progressive loss of coherence, loss of correlation holes, and increase of correlation squeezing.
Our findings further reinforce the uniqueness of sTG states in their stability and contextualize the effect of long-range interactions in a more realistic continuum description.

The rest of this paper is structured as follows.
In section \ref{sec:system}, we expostulate the theory behind the system we investigate and the quenching protocols we implement in our simulations.
In section \ref{sec:methods}, we briefly review the main features of our methods and we define the observables used to track the post-quench dynamics.
In sections \ref{sec:results-unit-filling} and \ref{sec:results-double-filling} we present our results for respectively unit and double filling and discuss their significance.
Finally, in section \ref{sec:conclusions}, we summarize the main findings of our paper and give an outlook on potential future research directions.
%%%%%%%%%%%%%%%%%%%%%%%%%%%%%%%%%%%%%%%%%%%%%%%%%%%%%%%%%%%%%%%%%%%%%%%%%%%%%%%%%%%%%%%%%%%%%

%%%%%%%%%%%%%%%%%%
%%%%% SYSTEM %%%%%
%%%%%%%%%%%%%%%%%%
%%%%%%%%%%%%%%%%%%%%%%%%%%%%%%%%%%%%%%%%%%%%%%%%%%%%%%%%%%%%%%%%%%%%%%%%%%%%%%%%%%%%%%%%%%%%%
\section{System and protocol}
\label{sec:system}
We are interested in the dynamics of a system of $N$ interacting bosons of mass $m$ in a one-dimensional lattice, which is governed by the time-dependent many-body Schr\"odinger equation 
\begin{equation}
\hat{H} \psi = i \hbar \frac{\partial \psi}{\partial t}.
\end{equation}
The total Hamiltonian has the form
\begin{equation} 
\hat{H}(x_1,x_2, \dots x_N)= \sum_{i=1}^{N} \hat{h}(x_i) + \sum_{i<j=1}^{N}\hat{W}(x_i - x_j).
\label{propagation_eq}
\end{equation}
Its one-body part is $\hat{h}(x) = \hat{T}(x) + \hat{V}_{OL}(x)$, where $\hat{T}(x) = -\frac{\hbar^2}{2m} \frac{\partial^2}{\partial x^2}$ is the kinetic energy operator and $\hat{V_{OL}}(x) = V_0 \sin^{2}(k_0 x)$ is the external 1D lattice of depth $V_0$ and wave vector $k_0$.
We remark that the Hamiltonian $\hat{H}$ can be written in dimensionless units obtained by dividing the dimensionful Hamiltonian by $\frac{\hbar^2}{mL^2}$, with $L$ an arbitrary length scale, which for the sake of our calculations we will set to be the period of the optical lattice.
Unless otherwise stated, we will keep $V_0$ fixed at 20 $E_r$, where $E_r = \frac{\hbar^2 k_0^2}{2 m}$ is the recoil energy of the lattice.
This depth is necessary to reach the Mott regime in the state preparation. 
We restrict the geometry to encompass $S$ sites in the optical lattice (defined as the spatial extent between two maxima in the sinusoidal function) adding hard-wall boundary conditions at each end.
The two-body interactions $\hat{W}(x_i - x_j)$ contain both short-range $W_C(x_i,x_j)= g_0 \delta(x_i - x_j)$ and long-range dipolar interactions $W_D(x_i,x_j) = \frac{g_d}{|x_i-x_j'|^3 + \alpha}$.
Here, $g_0$ is the strength of the local interactions, which is strongly positive before the quench and suddenly becomes negative after the quench, $g_d$ controls the strength of the long-range interactions and it is negative all throughout our calculations, and $\alpha$ denotes a renormalization factor used to avoid nonphysical singularities at $x_i=x_j$. 
Without long-range interactions, the ground state of the system is either in the TG Mott phase (with one particle per site) or in a multiple-filled Mott phase (with more than one particles per site). 
With strong negative DDIs, the ground state is a self-bound quantum droplet state at the central site in the optical lattice and possesses distinguished many-body features.
Weakly negative DDIs interpolate between these two cases, creating variations of cluster or droplet states where the central site has strong (potentially multiple) occupation but the neighboring sites see a rapid decrease in population.

With the Hamiltonian above, we will study two different setups.
First, we consider a unit filling scenario of $N=5$ particles in $S=5$ sites.
Then, we explore a more correlated double filling setup, with $N=6$ particles in $S=3$ sites.
The protocol in both cases is identical.
We first obtain the ground state for the Hamiltonian with strong contact repulsion ($g_0^{t<0} \gg 0$).
Then, at time $t=0$, we propagate this state by quenching it to the strongly attractive regime of equal magnitude ($g_0^{t \ge 0} \ll 0$, $|g_0^{t \ge 0}| = g_0^{t<0}$). 
In the experiments, this is achieved by exploiting a confinement-induced resonance (CIR), where particle interactions are enhanced in a confined space such as a trap or potential well~\cite{Olshanii:1998, Petrov:2000, Tiesinga:2000, Bergeman:2003, Kim:2005, Yurovsky:2005, Melezhik:2007, Saeidian:2008}. 
By means of Feshbach resonance techniques, it is possible to control the $s$-wave contact interaction and indirectly also control the long-range interactions.
We study the stability of the different initial states -- ranging from sTG to droplet -- after the sudden quench for different values of the attractive dipolar interactions $g_d$.
%%%%%%%%%%%%%%%%%%%%%%%%%%%%%%%%%%%%%%%%%%%%%%%%%%%%%%%%%%%%%%%%%%%%%%%%%%%%%%%%%%%%%%%%%%%%%

%%%%%%%%%%%%%%%%%%%
%%%%% METHODS %%%%%
%%%%%%%%%%%%%%%%%%%
%%%%%%%%%%%%%%%%%%%%%%%%%%%%%%%%%%%%%%%%%%%%%%%%%%%%%%%%%%%%%%%%%%%%%%%%%%%%%%%%%%%%%%%%%%%%%
\section{Methods}
\label{sec:methods}
To investigate the dynamics of dipolar bosons under the sudden quench, we employ the MultiConfigurational Time-Dependent Hartree method for indistinguishable particles (MCTDHB)~\cite{Streltsov:2006, Streltsov:2007, Alon:2007, Alon:2008} implemented by the MCTDH-X software~\cite{Alon:2008,Lode:2016,Fasshauer:2016,Lin:2020,Lode:2020,MCTDHX}.
MCTDH-X solves the many-body Schr\"{o}dinger equation by recasting the many-body wave function as an adaptive superposition of $M$ time-dependent permanents constructed from single-particle wave functions, called orbitals. 
Both the coefficients and the basis functions in this superposition are optimized in time to yield either ground-state information (via imaginary time propagation) or full-time dynamics (via real-time propagation).
We refer to the Appendix ~\ref{app:MCTDHX} for further details on this method. 

Utilizing several orbitals in the many-body ansatz allows us to capture fragmented many-body states, i.e. states that cannot be described by mean-field approximations, like sTG states.
In general, depending on the quench process and the quenched parameter of the initial Hamiltonian, fragmentation may increase (or less typically decrease) with time -- a phenomenon usually called dynamical fragmentation.
This prompts the question of orbital convergence for a fragmented state, i.e. how many orbitals are required to capture the dynamics correctly.
In MCTDHB, one method to ensure convergence is by repeating the computation with a higher number of orbitals until the dynamical measures converge and the occupation in the last orbital becomes insignificant. 
For very strong quenches, obtaining the exact, converged dynamics at long times might require a computationally unfeasible number of orbitals; therefore, only short and intermediate time dynamics can be obtained reliably.
For moderately strong quenches like the ones we are probing in this work, we are able to construct a converged dynamics during the entire time evolution by employing up to $M=15$ orbitals.
Details on the calculations that confirm the orbital convergence of our results are offered in appendix~\ref{app:orbital-conv}.

To extract information from the system, we calculate several observables from the many-body state $\left| \Psi(t) \right>$.
To probe the spatial distribution of the bosons, we calculate the one-body density as
\begin{align}
\rho(x; t) &= \left< \Psi(t) \right| \hat{\Psi}^{\dagger}(x) \hat{\Psi}(x) \left| \Psi(t) \right>.
\end{align}

To measure the degree of coherence and many-body correlation, we calculate the reduced one-body and two-body densities, defined respectively as 
\begin{align}
\rho^{(1)}(x,x'; t) &= \left< \Psi(t) \right| \hat{\Psi}^{\dagger}(x) \hat{\Psi}(x') \left| \Psi(t) \right> \\
\rho^{(2)}(x,x'; t) &= \left< \Psi(t) \right| \hat{\Psi}^{\dagger}(x) \hat{\Psi}^{\dagger}(x') \hat{\Psi}(x') \hat{\Psi}(x) \left| \Psi(t) \right>.
\end{align}
In appendix~\ref{app:obs}, we also showcase results pertaining to normalized one-body and two-body densities -- quantities normally referred to as one-body and two-body Glauber correlation functions $g^{(1)}(x,x') = \frac{\rho^{(1)}(x,x')}{N \sqrt{\rho(x) \rho(x')}}$ and $g^{(2)}(x,x')= \frac{\rho^{(2)}(x,x')}{N^2 \rho(x) \rho(x')}$. 

From the reduced one-body density matrix, it is possible to obtain information regarding the orbital occupation throughout the time evolution, i.e. how much of the provided Hilbert space is dynamically occupied.
This can be expressed via the natural orbitals $\phi_i^{(\mathrm{NO})}$ and the orbital occupations $\rho_i$, which are the eigenfunctions and eigenvalues of $\rho^{(1)}(x,x')$, i.e.
\begin{equation}
\rho^{(1)}(\mathbf{x},\mathbf{x}') = \sum_i \rho_i \phi^{(\mathrm{NO}),*}_i(\mathbf{x}')\phi^{(\mathrm{NO})}_i(\mathbf{x}).\label{eq:RDM1}
\end{equation}
The dynamics of the orbital occupations offer another instrument to quantify the stability (or lack thereof) of the initial state after the quantum quench. 

Finally, to study the eventual fate of all the different initial states -- from the TG Mott states to the single-site clusters -- and verify their stability under the sTG quench, we also calculate the autocorrelation function 
\begin{align}
    \mathcal{F}(t) \equiv |\left< \Psi(0) \middle| \Psi(t) \right>|^2.
\end{align}
This quantity offers an immediate, quantitative measure of the post-quench stability at the many-body level.
If it remains close to 1 for long times after the quench, we can consider the initial state stable with respect to the sudden quench.
A decay away from this value indicates instead a loss of coherence and the eventual destabilization of the initial state.
 
%%%%%%%%%%%%%%%%%%%%%%%%%%%%%%%%%%%%%%%%%%%%%%%%%%%%%%%%%%%%%%%%%%%%%%%%%%%%%%%%%%%%%%%%%%%%%

%%%%%%%%%%%%%%%%%%%
%%%%% RESULTS %%%%%
%%%%%%%%%%%%%%%%%%%
%%%%%%%%%%%%%%%%%%%%%%%%%%%%%%%%%%%%%%%%%%%%%%%%%%%%%%%%%%%%%%%%%%%%%%%%%%%%%%%%%%%%%%%%%%%%%
\section{Results for unit filling}
\label{sec:results-unit-filling}
We begin our analysis by presenting the results for the unit filling case of $N=5$ bosons in $S=5$ sites.
In this section, unless otherwise stated, the magnitude of the contact interactions is fixed at $|g_0| \approx 2 E_r$.
We have probed a fine array of dipolar interactions in the range $g_d \in [-0.4 E_r, 0 E_r]$, but we will only show representative results to be concise.

\subsection{Density}
%%%%%%%%%%%%%%%%%%%%%%%%%%%%%%%%%%%%%%%%%%%%
\begin{figure}
\centering
\includegraphics[width=1.0\columnwidth]{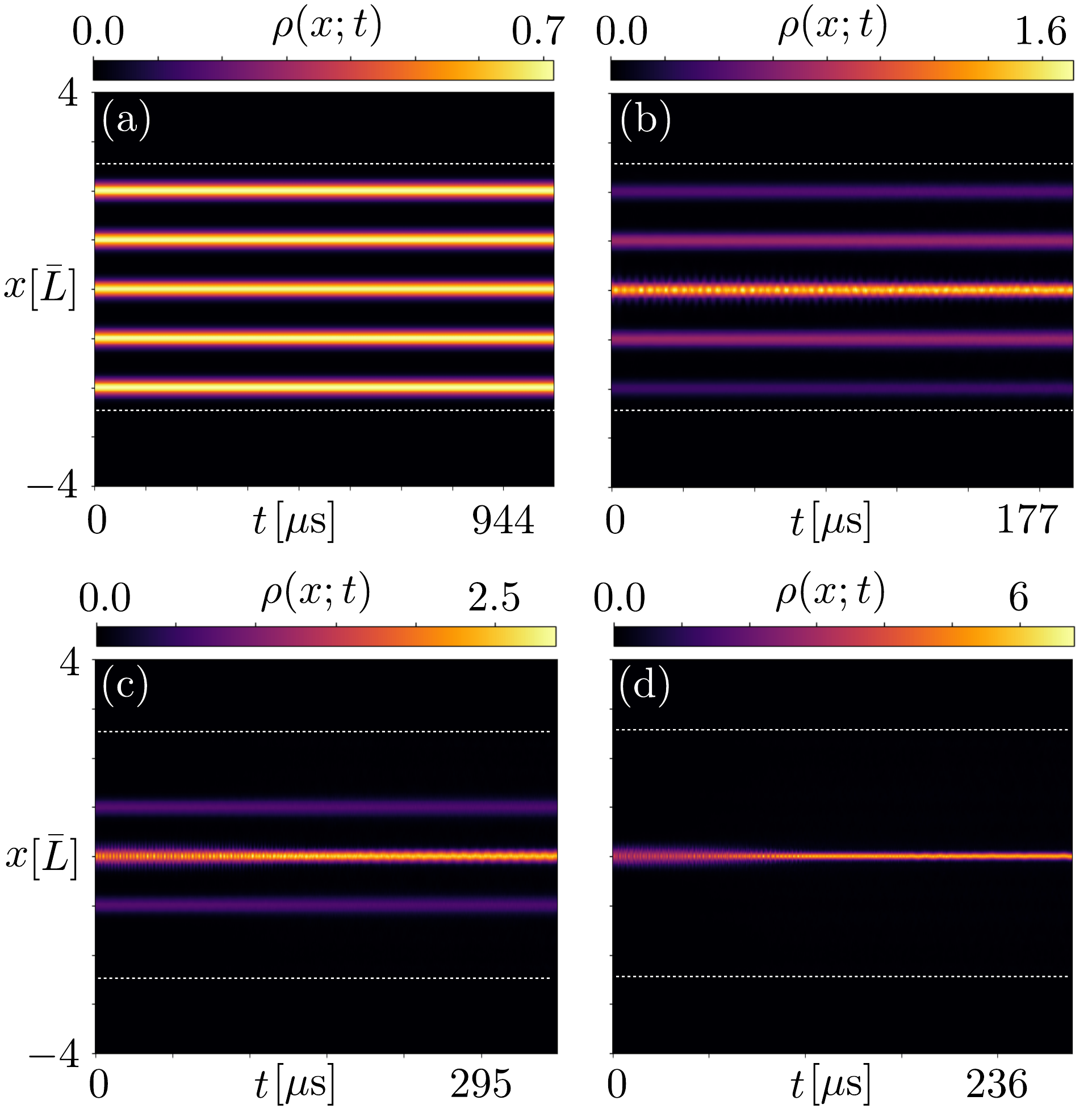}
\caption{
Density dynamics post quench for $N=5$ bosons and $M=15$ orbitals in $S=5$ sites (unit filling) with 
(a) $g_d=-0.16 E_r$, 
(b) $g_d=-0.18 E_r$,
(c) $g_d=-0.19 E_r$, 
(d) $g_d=-0.4 E_r$.
The other parameters are $V_0 \approx 20 E_r$ and $|g_0| \approx 2 E_r$.
The dotted lines indicate the system boundaries.
}
\label{fig:dens-single}
\end{figure}
%%%%%%%%%%%%%%%%%%%%%%%%%%%%%%%%%%%%%%%%%%%%

The density dynamics is plotted in Fig.~\ref{fig:dens-single}(a)-(d) for four representative values of $g_d$ showcasing the appearance of different dynamical behaviors. 
All computations shown in the figures have been performed with $M=15$ orbitals. 
We have systematically checked convergence in appendix~\ref{app:orbital-conv} from $M=12$ up to $M=15$ orbitals both for the initial states and for the time evolution, and determined that the occupation of the last few orbitals is negligible.
This confirms that our results are converged in the number of orbitals throughout the dynamics.

The initial state retains a fully fragmented, clean TG Mott state configuration for all $g_d$ values down to around $g_d=-0.16 E_r$.
This state has positive energy and remains highly stable for very long time after the quench.
Fig.~\ref{fig:dens-single}(a) shows the density dynamics for the TG to sTG quench in this case.
From a density perspective, we can appreciate the robustness of this state, which appears unaffected by the interaction quench.

At around $gd=-0.18 E_r$, the initial density profile begins to morph into a cluster state, which can have non zero population on all five sites [Fig.~\ref{fig:dens-single}(b), $gd=-0.18 E_r$] or only on the center-most three [Fig.~\ref{fig:dens-single}(c), $gd=-0.19 E_r$].
The system still has positive energy, but -- due to the effect of long-range interactions -- the five bosons are not uniformly localized.
Upon quenching, these cluster states overall retain their initial shape, but oscillations in the density of each peak and a slow but progressive accumulation of the density towards the center can be observed.
This already signals a destabilization with respect to the initial configuration, as opposed to the pure TG Mott state.
We remark that the parameter space for the observation of these intermediate cluster states is very narrow, in the order of 0.02 $E_r$.
This is reminiscent of the tiny sliver in parameter space where a quantum liquid phase can be observed in extended Bose-Hubbard models~\cite{Marciniak:2023}.

For even stronger attractions $g_d < -0.2 E_r$, the bosons form a self-bound droplet with negative energy in the central lattice. 
Upon quenching, the collapse of the initial state is much more pronounced.
In the order of $\approx 70$ $\mu$s, the initial droplet is compressed to a much narrow range due to the strong attractions.
This is illustrated in Fig.~\ref{fig:dens-single}(d) for $g_d= -0.4 E_r$.

%%%%%%%%%%%%%%%%%%%%%%%%%%%%%%%%%%%%%%%%%%%%
\begin{figure}
\centering
\includegraphics[width=\columnwidth]{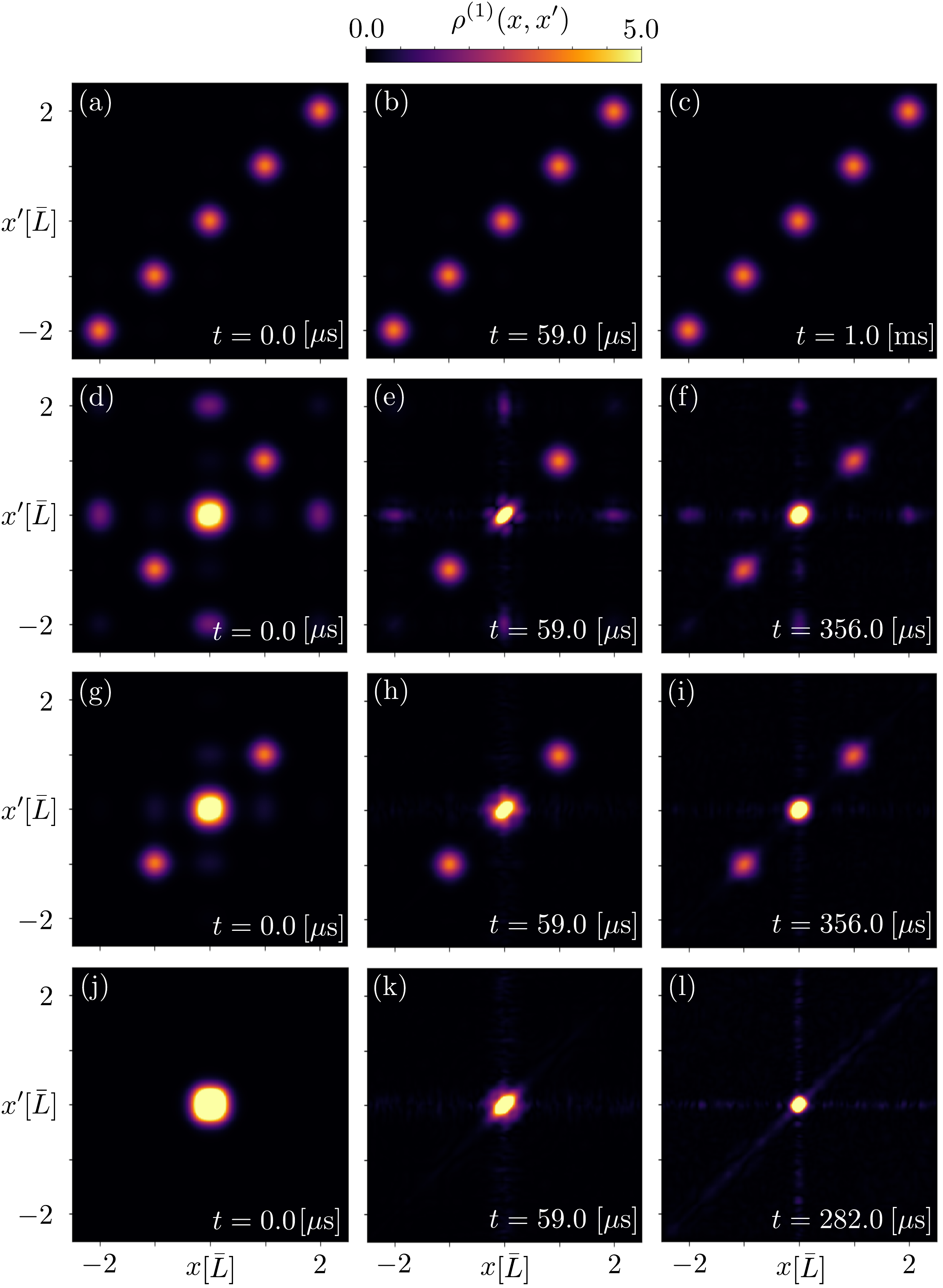}
\caption{
Dynamics of the reduced one-body density $\rho^{(1)}(x,x')$ post quench for $N=5$ bosons and $M=15$ orbitals in $S=5$ sites (unit filling) with 
(a)-(c) $g_d=-0.16 E_r$, 
(d)-(f) $g_d=-0.18 E_r$,
(g)-(i) $g_d=-0.19 E_r$, 
(j)-(l) $g_d=-0.4 E_r$.
The other parameters are $V_0 \approx 20 E_r$ and $|g_0| \approx 2 E_r$.
}
\label{fig:rho1-single}
\end{figure}
%%%%%%%%%%%%%%%%%%%%%%%%%%%%%%%%%%%%%%%%%%%%

%%%%%%%%%%%%%%%%%%%%%%%%%%%%%%%%%%%%%%%%%%%%
\begin{figure}
\centering
\includegraphics[width=\columnwidth]{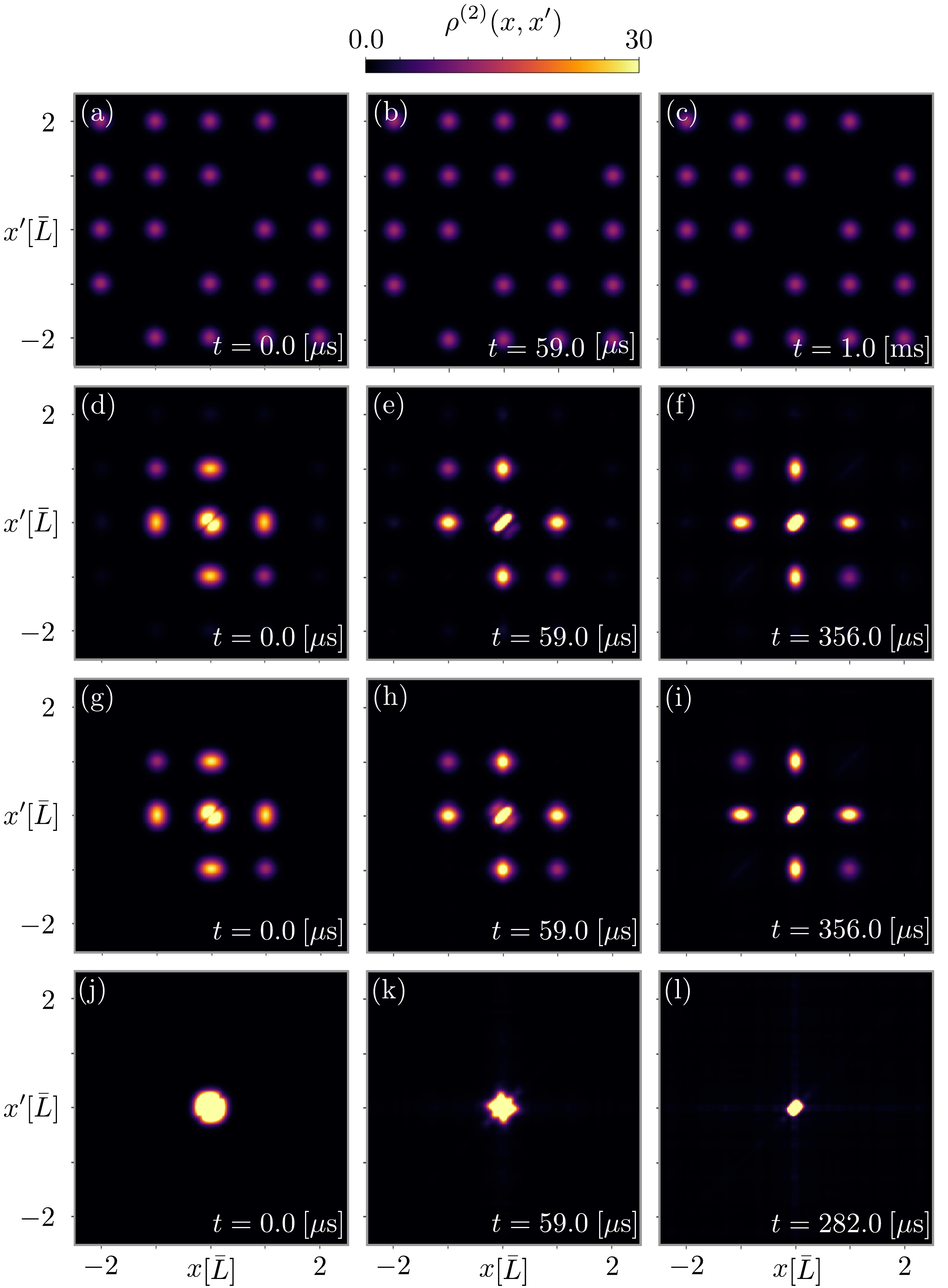}
\caption{
Dynamics of the reduced two-body density $\rho^{(2)}(x,x')$ post quench for $N=5$ bosons and $M=15$ orbitals in $S=5$ sites (unit filling) with 
(a)-(c) $g_d=-0.16 E_r$,
(d)-(f) $g_d=-0.18 E_r$,
(g)-(i) $g_d=-0.19 E_r$, 
(j)-(l) $g_d=-0.4 E_r$.
The other parameters are $V_0 \approx 20 E_r$ and $|g_0| \approx 2 E_r$.
}
\label{fig:rho2-single}
\end{figure}
%%%%%%%%%%%%%%%%%%%%%%%%%%%%%%%%%%%%%%%%%%%%

\subsection{Correlations}
To clearly explore the many-body features in the sTG quench, we present snapshots of the reduced one-body density $\rho^{(1)}(x,x^{\prime},t)$ and two-body density $\rho^{(2)}(x,x^{\prime},t)$ for some chosen times in Fig.~\ref{fig:rho1-single} and Fig.~\ref{fig:rho2-single}, respectively.
These quantities illustrate coherence and correlation effects at the one-body and two-body level.

For the TG Mott states with long-range attractions down to $g_d=-0.16 E_r$, the one-body density is perfectly diagonal, indicating only self-correlation within each Mott peak.
The two-body density, exhibits a completely depleted region along the diagonal -- termed ``correlation hole'' --  which is a manifestation of fermion-like exclusion stemming from the strong repulsion that makes the bosons act as hard-core particles.
As already seen from the density, the TG to sTG quench at zero or weak dipolar interactions [Figs.~\ref{fig:rho1-single}(a)-(c) and \ref{fig:rho2-single}(a)-(c)] is stable for very long times, and both the diagonal shape of $\rho^{(1)}$ and the correlation hole of $\rho^{(2)}$ persist at all probed times.

However, further increasing the dipolar interactions, e.g. to $g_d=-0.18 E_r$, leads to a transition from the TG Mott state to a progressively more localized cluster state, which is apparent also in the correlation functions [Figs.~\ref{fig:rho1-single}(d)-(f) and \ref{fig:rho2-single}(d)-(f)].
In $\rho^{(1)}$, we can distinguish the existence of coherence across the different sites at $t=0$. 
After the quench, the off-diagonal coherence progressively decreases and almost disappears at time $t=356$ $\mu$s.
The diagonal coherence is also visibly squeezed within each site, indicating a change in the on-site correlation structure.
A similar fate is observed for $\rho^{(2)}$, in particular with respect to the correlation hole that disappears.
For $gd=-0.19 E_r$, the correlation with the outer sites has completely vanished from the start.
During time evolution, we can see how the diagonal correlation becomes progressively more localized under the influence of the attractive interactions [Figs.~\ref{fig:rho1-single}(g)-(i) and \ref{fig:rho2-single}(g)-(i)]. 

For yet stronger dipolar interactions, the initial state is a highly localized droplet with nonzero correlations only within the central site.
Upon quenching, the correlations are rapidly squeezed to an even narrower range, both in the one-body and two-body reduced density matrix [Figs.~\ref{fig:rho1-single}(j)-(l) and \ref{fig:rho2-single}(j)-(k)]. 

In summary, the above analysis of the post quench correlation dynamics shows that when dipolar interactions are strong enough to perturb the initial state away from the pure TG Mott configuration, a collapse under the sTG quench procedure ultimately occurs.
On the contrary, the pure and weakly dipolar TG Mott state does not exhibit intersite correlation and can retain its overall correlation pattern despite the contact interaction quench.

\subsection{Autocorrelation function}
The density, one-body, and two-body reduced density matrices analyzed so far indicate that the stability of the initial state is retained only under weak dipolar attractions that minimally perturb the TG Mott configuration. 
However, these quantities provide only few-body information on the system's time evolution. 
It remains unclear whether the apparent stability observed in the density for the sTG state persists at the full many-body level. 
Is there indeed a threshold in $g_d$ below which the sTG state remains unperturbed?

To fully characterize the many-body dynamics, a many-body observable is essential. 
One such observable is the autocorrelation function $\mathcal{F}(t)$, introduced earlier, which measures the overlap or fidelity between the many-body state at time $t$ and its initial configuration. 
This function provides a direct indication of any deviation from stability during time evolution.

\begin{figure}
\centering
\includegraphics[width=\columnwidth]{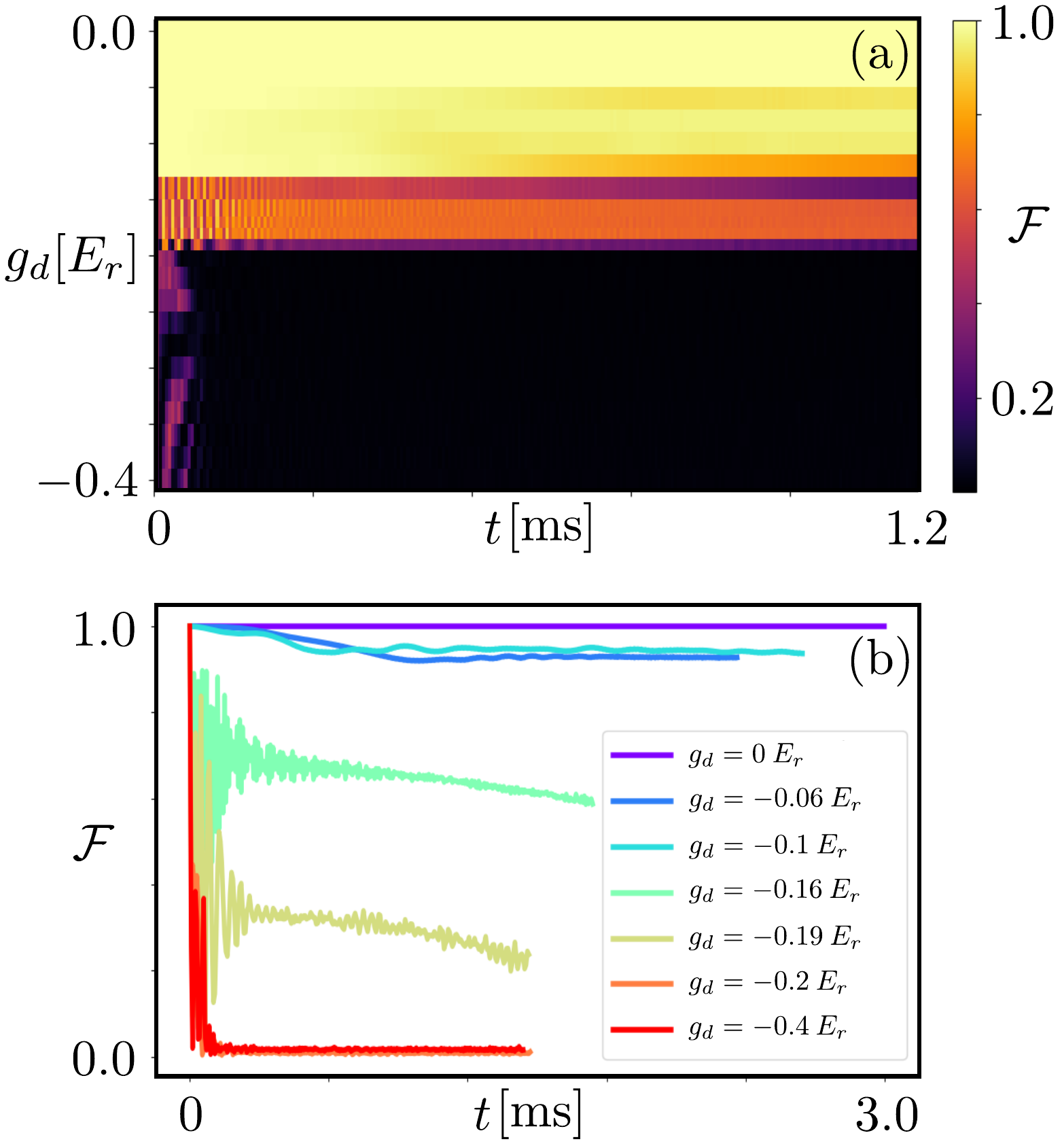}
\caption{
Dynamics of the autocorrelation function $\mathcal{F}(t)$ post quench for $N=5$ bosons and $M=12$ orbitals in $S=5$ sites (unit filling).
(a) Dynamics of $\mathcal{F}(t)$ in the entire range of probed dipolar attractions $g_d$ from $0.0$ to $-0.4 E_r$.
(b) Dynamics of $\mathcal{F}(t)$ for a few selected values of $g_d$.
The other parameters are $V_0 \approx 20 E_r$ and $|g_0| \approx 2 E_r$.
}
\label{fig:autocorr-single}
\end{figure}

Figure~\ref{fig:autocorr-single} presents the results for the autocorrelation function in the unit-filling setting. 
To probe the longest possible times with our computational resources, we show results obtained with $M=12$ orbitals. 
In Appendix~\ref{app:autocorr}, we also present results for $M=15$ orbitals, which are nearly indistinguishable from those shown here at shorter times, suggesting that our autocorrelation function is converged within the probed timescales.

From Fig.~\ref{fig:autocorr-single}(a), we identify three distinct regimes. 
The first regime, characterized by high fidelity throughout time evolution, shows that the initial state is only weakly perturbed at the many-body level. 
This region spans values of $g_d$ from 0 to around $-0.1 E_r$.
Surprisingly, this is a much narrower range than the density stability would suggest. 
In the cuts presented in Fig.~\ref{fig:autocorr-single}(b), we even observe that perfect autocorrelation is preserved only in complete absence of dipolar interactions. 
Even minimal dipolar attractions, such as $g_d \approx -0.06 E_r$, eventually reduce the fidelity to around 0.9 at longer times.

The second regime shows significantly reduced stability during dynamics, marked in orange and red shades in Fig.~\ref{fig:autocorr-single}(a), corresponding to intermediate values of $g_d$ in the interval $[-0.2 E_r, -0.1 E_r]$. 
Remarkably, this interval includes both TG Mott states and cluster states. 
These states exhibit a rapid fidelity decrease during dynamics, with values reaching between $\mathcal{F} \approx 0.3 - 0.7$ as shown in Fig.~\ref{fig:autocorr-single}(b), indicating that the quench strongly affects the state at the many-body level.

Finally, the third regime features a rapid decrease in the autocorrelation function to near-zero values, primarily involving highly localized cluster states and droplets with $g_d < -0.19 E_r$. 
The collapse of fidelity to zero reflects a complete reorganization of the many-body state following the quench, driven by strong dipolar attractions and quenched contact interactions.

%%%%%%%%%%%%%%%%%%%%%%%%%%%%%%%%%%%%%%%%%%%%
%%%%%%%%%%%%%%%%%%%%%%%%%%%%%%%%%%%%%%%%%%%%

%%%%%%%%%%%%%%%%%%%%%%%%%%%%%%%%%%%%%%%%%%%%%%%%%%%%%%%%%%%%%%%%%%%%%%%%%%%%%%%%%%%%%%%%%%%%%
\section{Results for Double filling}
\label{sec:results-double-filling}
We now present results for more exotic doubly-filled initial states consisting of $N$=6 bosons in $S=3$ sites.
Unless otherwise stated, we keep the magnitude of the contact interactions at $|g_0| \approx 2 E_r$ and probe the same range of dipolar interactions $g_d \in [-0.4 E_r, 0 E_r]$ as in the single filling case, showing a handful of representative cases.

%%%%%%%%%%%%%%%%%%%%%%%%%%%%%%%%%%%%%%%%%%%%
\begin{figure}
\centering
\includegraphics[width=\columnwidth]{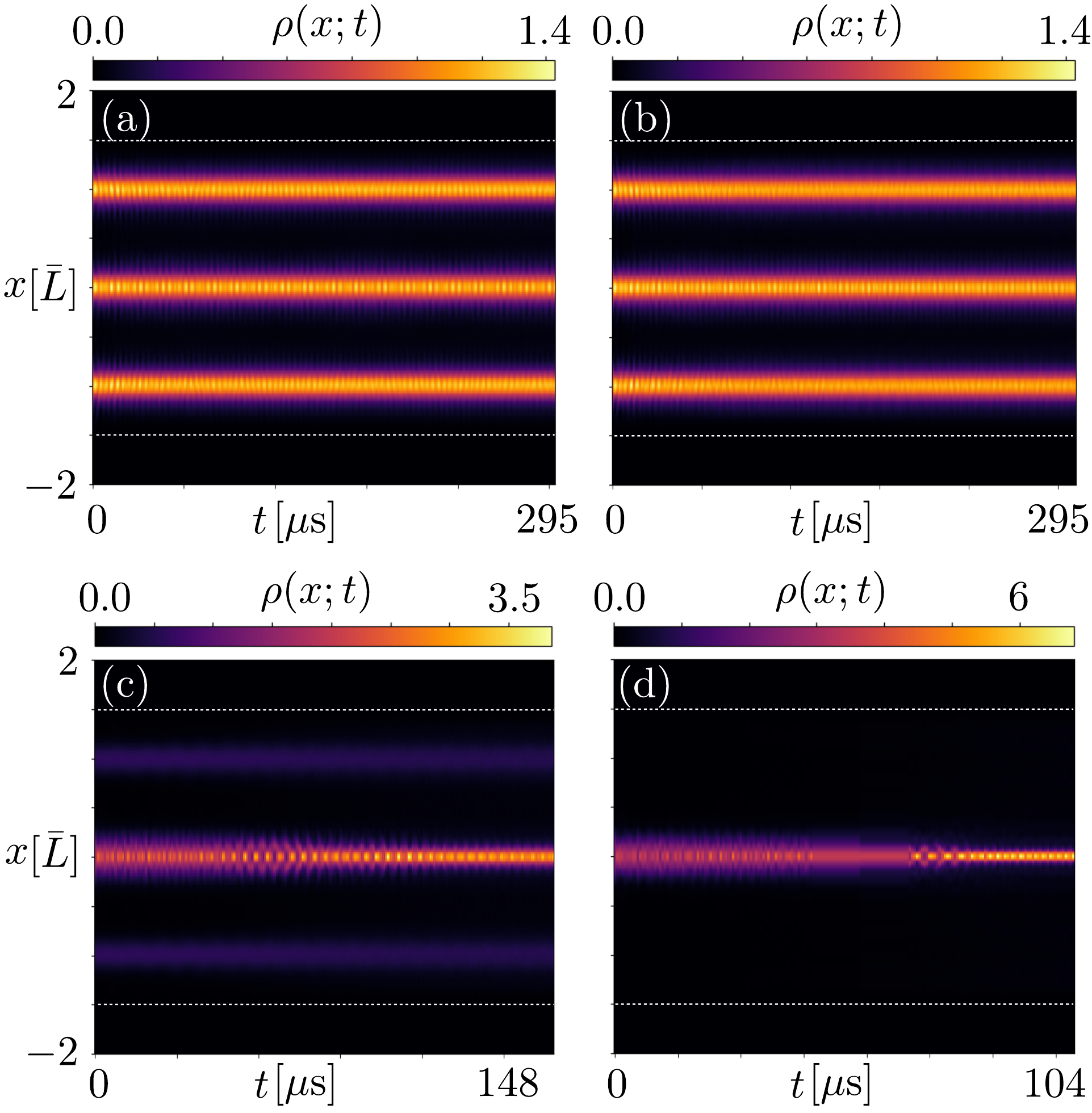}
\caption{
Density dynamics post quench for $N=6$ bosons and $M=15$ orbitals in $S=3$ sites (double filling) with 
(a) $g_d=-0.16 E_r$, 
(b) $g_d=-0.17 E_r$, 
(c) $g_d=-0.19 E_r$.
(d) $g_d=-0.4 E_r$.
The other parameters are $V_0 \approx 20 E_r$ and $|g_0| \approx 2 E_r$.
The dotted lines indicate the system boundaries.
}
\label{fig:dens-double}
\end{figure}
%%%%%%%%%%%%%%%%%%%%%%%%%%%%%%%%%%%%%%%%%%%%

\subsection{Density}
Fig.~\ref{fig:dens-double} illustrates the post-quench density dynamics for increasing values of dipolar interaction strength $g_d$.
For all the values probed, we find that the system is rather unstable to the sudden quench and the density rapidly exhibits oscillatory patterns and/or collapse under the attractive forces.
For $g_d=0.0 E_r$ -- and all the way down to $g_d=-0.16 E_r$ -- the initial state consists of three pairs of bosons located at the minima of the optical lattice, i.e. a double-filled Mott state.
Upon quenching, the three-peak structure is retained for long times.
However, from Fig.~\ref{fig:dens-double}(a) we can clearly see oscillatory patterns in each peak, signalling that the double-filled Mott state is not robust to the sudden quench perturbation.
This is in stark contrast to the unit filling case, where the TG Mott state -- despite showing instability at the many-body level -- still exhibits extreme stability in the density even down to values of $g_d=-0.18 E_r$.
We thus reveal that the stability of the sTG state does not survive multiple filling, even at $g_d=0.0$.

When the dipolar attractions are increased, the initial state becomes more clustered. 
This initially appears at around $g_d=-0.17 E_r$, where the central peak absorbs slightly more density than the two side peaks.
This is barely visible in Fig.~\ref{fig:dens-double}(b), but it can be appreciated in the initial density figure in appendix~\ref{app:orbital-conv}.
When the initial density only slightly deviates from the double-filled Mott state, the dynamics is essentially the same: the three-peak structure is preserved albeit with small on-site oscillations.
By further increasing the magnitude of the dipolar interactions to $g_d=-0.19 E_r$, however, we promptly achieve a much more localized cluster state.
This cluster state is even more sensitive to the density perturbations caused by the interaction quench, as we can see from the more pronounced oscillations coupled with a narrowing of the peaks in Fig.~\ref{fig:dens-double}(c).

Finally, when the long-range interactions reach $g_d = -0.2 E_r$, we achieve an initial single-peak state.
Similarly to the fate of the single-site droplet in the unit filling case, this state is quite unstable to quenching.
As shown in Fig.~\ref{fig:dens-double}(d), it rapidly undergoes density oscillations that culminate in a much more compact density squeezed by the strong attraction.
We also remark that the time scales for this collapse are much faster, in the order of 100 $\mu$s.

%%%%%%%%%%%%%%%%%%%%%%%%%%%%%%%%%%%%%%%%%%%%
\begin{figure}
\centering
\includegraphics[width=\columnwidth]{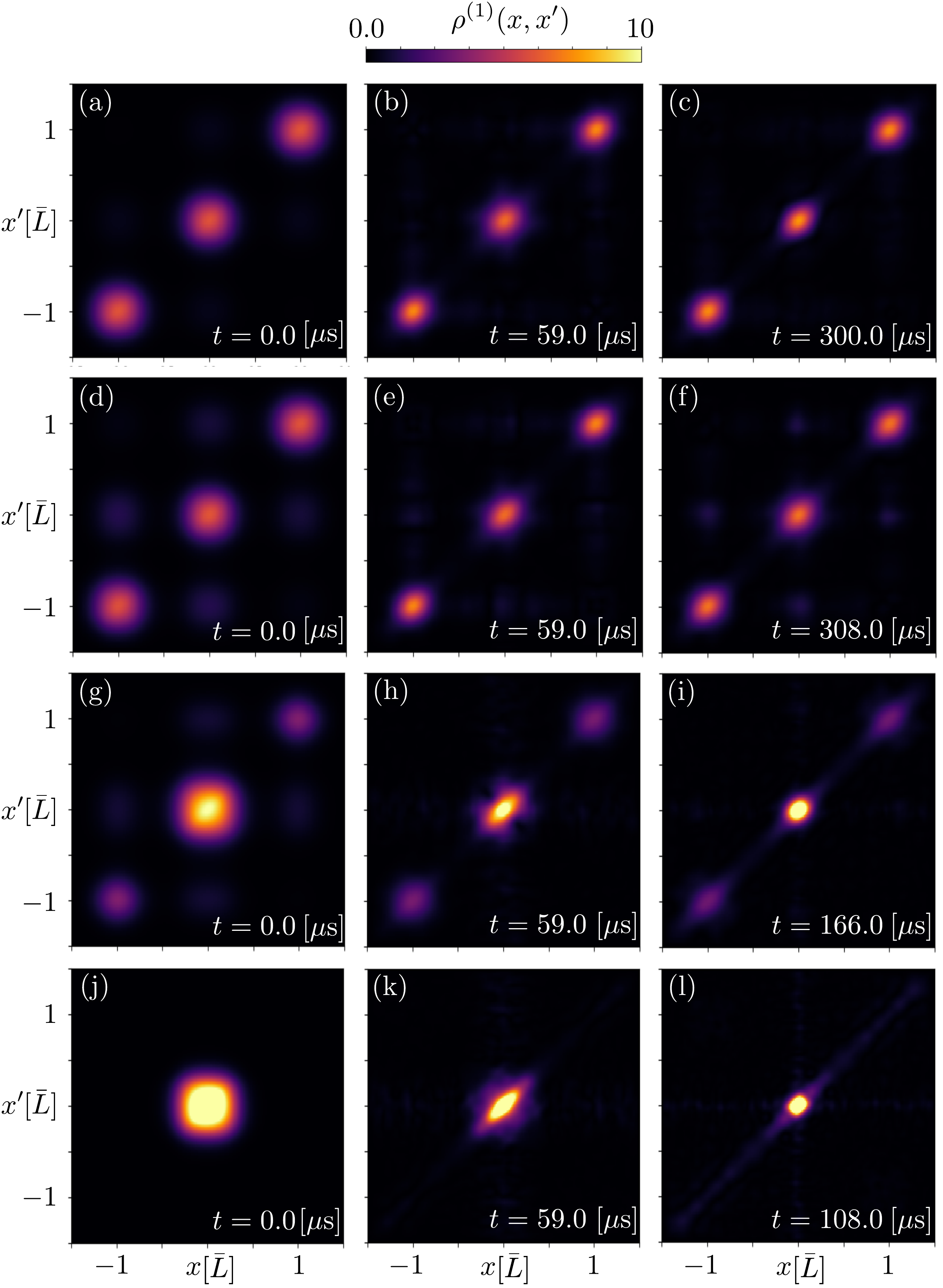}
\caption{
Dynamics of the reduced one-body density $\rho^{(1)}(x,x')$ post quench for $N=6$ bosons and $M=15$ orbitals in $S=3$ sites (double filling) with 
(a)-(c) $g_d=-0.16 E_r$, 
(d)-(f) $g_d=-0.17 E_r$, 
(g)-(i) $g_d=-0.19 E_r$,
(j)-(l)  $g_d=-0.4 E_r$.
The other parameters are $V_0 \approx 20 E_r$ and $|g_0| \approx 2 E_r$.
}
\label{fig:rho1-double}
\end{figure}
%%%%%%%%%%%%%%%%%%%%%%%%%%%%%%%%%%%%%%%%%%%%

%%%%%%%%%%%%%%%%%%%%%%%%%%%%%%%%%%%%%%%%%%%%
\begin{figure}
\centering
\includegraphics[width=\columnwidth]{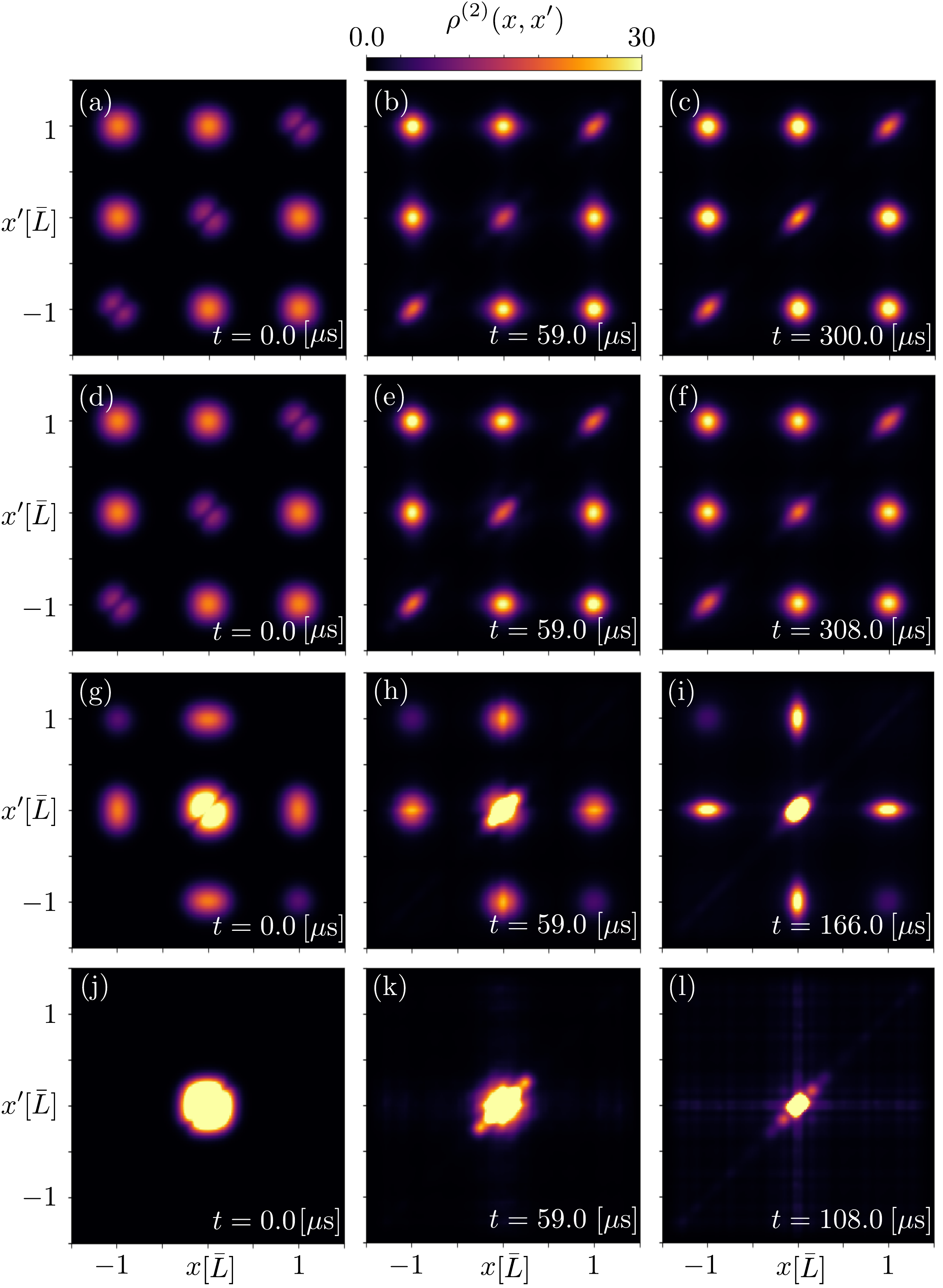}
\caption{
Dynamics of the reduced two-body density $\rho^{(2)}(x,x')$ post quench for $N=6$ bosons and $M=15$ orbitals in $S=3$ sites (double filling) with 
(a)-(c) $g_d=-0.16 E_r$, 
(d)-(f) $g_d=-0.17 E_r$, 
(g)-(i) $g_d=-0.19 E_r$,
(j)-(l)  $g_d=-0.4 E_r$.
The other parameters are $V_0 \approx 20 E_r$ and $|g_0| \approx 2 E_r$.
Note that the colorbar is chosen in a range that makes fainter off-diagonal features more visible.
}
\label{fig:rho2-double}
\end{figure}
%%%%%%%%%%%%%%%%%%%%%%%%%%%%%%%%%%%%%%%%%%%%

\subsection{Correlations}

The mechanism for the collapse process at different dipolar interaction strengths becomes again clearer upon examining the one-body and two-body reduced density matrices visualized in Figs.~\ref{fig:rho1-double} and \ref{fig:rho2-double}.
For the double-filled Mott initial state (both at zero and non-zero $g_d$), the bosons display only diagonal one-body correlation throughout the time evolution, with the correlation becoming progressively more confined under the attraction as time goes on [Figs.~\ref{fig:rho1-double}(a)-(c)].
In the two-body reduced density matrix, the initial state clearly reveals two bosons per site, which are separated by a small correlation hole.
However, this correlation hole quickly vanishes under the influence of the attractive forces, merging the two correlation blobs along the diagonal into a single squeezed one [Figs.~\ref{fig:rho2-double}(a)-(c)].
The situation is very similar once we deviate slightly from the homogeneous double-filled case at $g_d=-0.17 E_r$ [Figs.~\ref{fig:rho2-double}(d)-(f)].
However, in this case a faint trace of off-diagonal one-body correlation can be identified in the initial state, which seems to persist for longer times [Figs.~\ref{fig:rho1-double}(d)-(f)].

For more prominent cluster states, e.g. at $g_d=-0.19 E_r$, the off-diagonal one-body coherence at time zero is quickly dissipated after the quench [Figs.~\ref{fig:rho1-double}(g)-(i)], while the diagonal correlation is strongly squeezed.
In the two-body reduced density matrix, we observe a similar dissipation of off-diagonal correlation between the two side-peaks, while the correlation hole is rapidly populated and the overall correlation is squeezed to narrower regions in the central peak [Figs.~\ref{fig:rho2-double}(g)-(i)].

Finally, below $g_d=-0.2 E_r$ we enter the droplet regime where only one peak is populated.
As already observed for the density, this state is highly susceptible to the quantum quench.
Both the one-body [Figs.~\ref{fig:rho1-double}(j)-(l)] and the two-body [Figs.~\ref{fig:rho2-double}(j)-(l)] correlation is rapidly confined to smaller areas as the strong attractions overcome the more spread-out initial state.

\subsubsection{Autocorrelation function}

\begin{figure}
\centering
\includegraphics[width=\columnwidth]{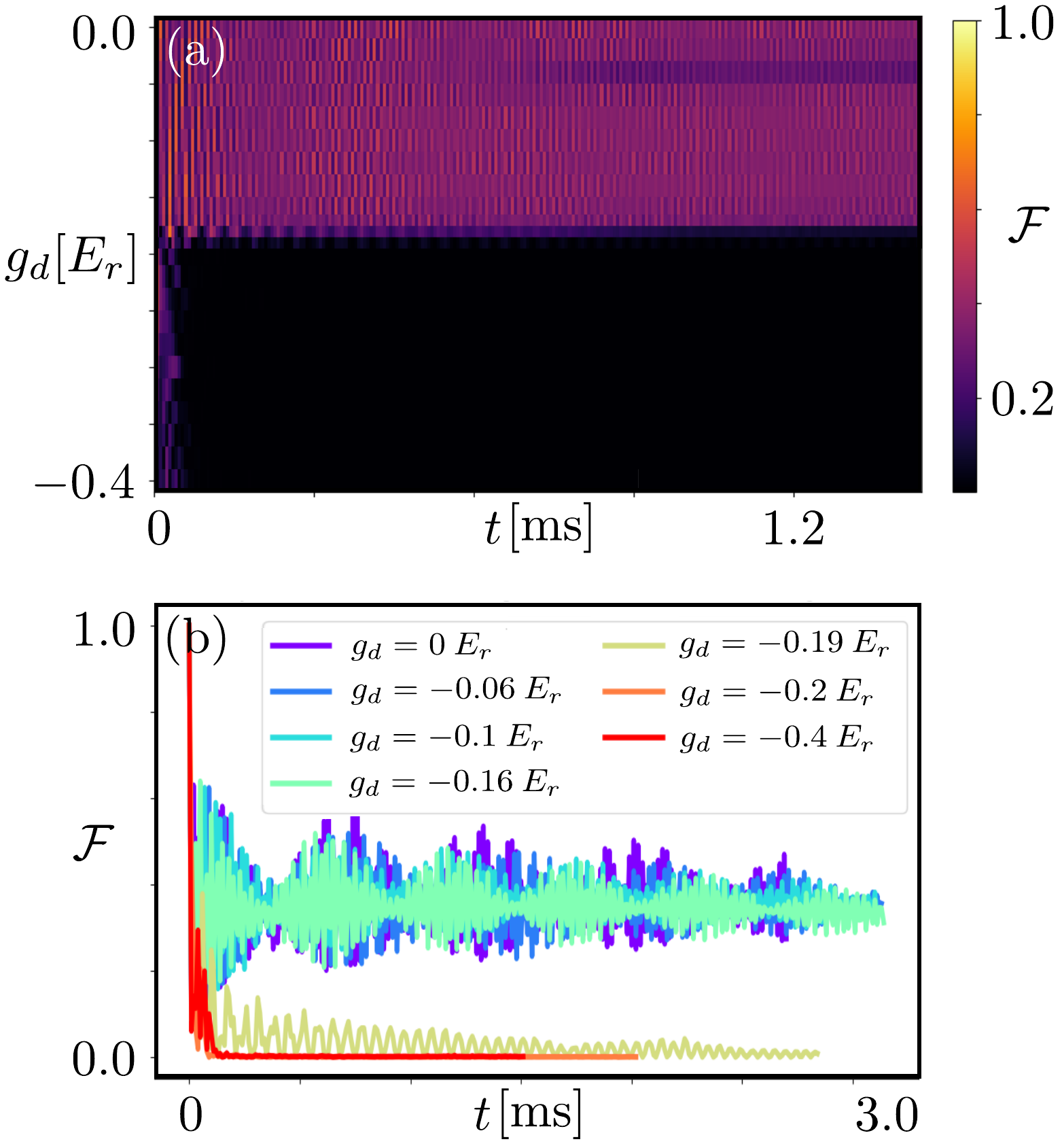}
\caption{
Dynamics of the autocorrelation function $\mathcal{F}(t)$ post quench for $N=6$ bosons and $M=12$ orbitals in $S=3$ sites (double filling).
(a) Dynamics of $\mathcal{F}(t)$ in the entire range of probed dipolar attractions $g_d$ from $0.0$ to $-0.4 E_r$.
(b) Dynamics of $\mathcal{F}(t)$ for a few selected values of $g_d$.
The other parameters are $V_0 \approx 20 E_r$ and $|g_0| \approx 2 E_r$.
}
\label{fig:autocorr-double}
\end{figure}

The stronger instability with respect to the interaction quench exhibited by the double-filled states can be better quantified by looking at the autocorrelation function.
This is shown in Fig.~\ref{fig:autocorr-double} again in two ways: in panel (a) by showing the time evolution of $\mathcal{F}(t)$ across all values of $g_d$, and in panel (b) for a few selected cuts.
In this case, the high-stability region has completely disappeared and only the two cases of low stability (purple region in Fig.~\ref{fig:autocorr-double}(a)) and complete many-body disruption (black region in Fig.~\ref{fig:autocorr-double}(a)) remain.
In fact, all values of $g_d$ between 0 and $-0.19 E_r$ display a rapid decay to a low saturation value of the fidelity around 0.4 with some beating patterns.
For even strong dipolar attraction, the autocorrelation function drops instead all the way down to zero, indicating a complete change of the many-body state.

A comparison between single-filling and double-filling systems highlights the critical role of coherence in the stability of the dynamics. 
In the single-filling case, the loss of off-diagonal coherences significantly contributes to the decay of the initial state. 
This effect is particularly evident in the sensitivity of the fidelity to the dipolar interaction strength in the region $g_d > -0.2 E_r$ (see Fig.~\ref{fig:autocorr-single}).
Conversely, in the double-filling case, the dynamics of the fidelity remains largely unchanged within the same $g_d > -0.2 E_r$, which could be attributed the attractive dipolar interactions suppressing off-diagonal coherences.
The difference between the two filling scenarios underscores the importance of coherence in determining the stability of the many-body state across different state configurations.

%%%%%%%%%%%%%%%%%%%%%%%%%%%%%%%%%%%%%%%%%%%%
%%%%%%%%%%%%%%%%%%%%%%%%%%%%%%%%%%%%%%%%%%%%

%%%%%%%%%%%%%%%%%%%%%%%%%%%%%%%%%%%%%%%%%%%%%%%%%%%%%%%%%%%%%%%%%%%%%%%%%%%%%%%%%%%%%%%%%%%%%

%%%%%%%%%%%%%%%%%%%%%%%
%%%%% CONCLUSIONS %%%%%
%%%%%%%%%%%%%%%%%%%%%%%
%%%%%%%%%%%%%%%%%%%%%%%%%%%%%%%%%%%%%%%%%%%%%%%%%%%%%%%%%%%%%%%%%%%%%%%%%%%%%%%%%%%%%%%%%%%%%
\section{Conclusions} 
\label{sec:conclusions}
We have investigated the quench dynamics of strongly-interacting bosons in an optical lattice for a wide range of dipolar interaction strengths and different fillings, generalizing the sTG quench procedure to the presence of long-range interactions and more complex initial states.
Our protocol consisted in preparing an initial state of bosons with moderate contact repulsions and weaker dipolar attractions, and then quenching the strength of the contact interactions to the negative (attractive) regime.
We have calculated numerous observables to study the time evolution, including one-body density, reduced one-body and two-body densities, natural occupations, and autocorrelation functions.
By analyzing these quantities, we have mapped out the effect of different dipolar interaction strengths on the stability of the post-quench states.

As expected, starting from a unit-filled Mott state at zero dipolar interactions we recollect the robustness of the original sTG quench, where the localized Mott peaks remain unperturbed for a very long time.
At weak dipolar interactions, this stability is also retained for comparably long times, as the Mott peaks remains localized and uncorrelated. 
By increasing the dipolar interactions further, however, the initial bosonic configuration progressively condenses into localized clusters.
The cluster states are much more sensitive to the quench procedure and a calculation of the autocorrelation function reveals a rapid loss of fidelity after the quench.
A similar fate is observed by starting from doubly filled states, be them in doubly-occupied Mott configurations (lower dipolar interactions) or cluster states and solitons (higher dipolar interactions).
The state rapidly collapses under the action of the combined attractions.
Overall, we found no evidence of intermediate prethermal states in the dynamics.

By examining different correlations measures, we find that the cause for this behavior lies within the structure of the initial states, that exhibit pockets of correlations within the dimers or within different peaks of the cluster states.
These pockets of correlations spread upon quenching and lead to instability and eventual collapse.
In particular, the comparison between single- and double-filling systems highlights the critical role of coherence in stabilizing dynamics, with the loss of off-diagonal coherences driving sensitivity to interaction strength in the single-filling case, while being suppressed in the double-filling case due to attractive dipolar interactions. 
This underscores the importance of coherence in determining the stability of many-body states.

Our study sheds light onto the behavior and stability of strongly correlated bosonic states in the presence of long-ranged interactions.
It also showcases the potential of harnessing such long-range interactions and quenching protocols to achieve novel pathways to the creation and control of excited quantum states of matter.
These questions are particularly timely since the systems we have investigated should be easily engineered in state-of-the-art ultracold experiments, e.g. with magnetic quantum gases~\cite{Chomaz:2023} or dipolar molecules that can be controlled with unprecedented precision~\cite{Bigagli:2023}. 
The complex atomic structure and the large dipole moment of these systems facilitate the control of the interatomic interaction strength, specially the relative strength of contact over dipolar interactions. 
This will lead to further exploration of few and many-body physics. 

In the future, we envision that similar approaches could be employed to study the complex interplay of different energy scales introduced by the competition between localizing optical lattices, local (contact) interactions, and long-range interactions, which could lead to a multitude of exotic quantum states.
Our study could also be extended to analyze the stability of other fermionization and crystallization phenomena, potentially in higher dimensions, in disordered landscapes, or fostered by other kind of long-range interactions, e.g. mediated by cavities~\cite{Molignini:2022, Lukas:2023}.
This last setup is particularly promising as it has been shown to give rise to a plethora of intriguing equilibrium and out-of-equilibrium phases of matter, such as higher-order superfluidity, Mott phases, and limit cycles~\cite{Molignini:2018,Lin:2019,Lin:2020-PRA,Lin:2021}.
It would be interesting to analyze whether the interplay of such phases with dipolar interactions could lead to a stabilization of prethermal states.

Another interesting route to pursue is the connection with the accuracy of dipolar quantum simulators~\cite{Hughes:2023}.
Our method solves the many-body Schr\"{o}dinger equation in the continuum, and thus should be an accurate representation of the physics at play in the actual quantum simulators and not just in the simulated lattice models (e.g. extended Bose-Hubbard models) studied elsewhere in the literature.
For example, a relevant question to ask is to what extend a similar interaction quench would lead the physics to deviate from that of a lattice model as a function of dipolar interaction strength, optical lattice depth, and other system parameters. 
It could also include the investigation of quantum many-body scar states which can give rise to infinitely long-lived coherent dynamics under quantum quenches from well controlled initial states~\cite{Sanjay:2022}. 

Another open avenue for future studies is the impact of long-range interactions on quantum holonomy -- where parameters are tuned to realize a cycle starting from the free system, to Tonks-Girardeau and super-Tonks-Girardeau regimes, and then back to the free system~\cite{Cheon:2013}. 
The comparison between sudden quench and adiabatic quench will help to understand the underlying correlation dynamics.
Furthermore, continuous quenches could provide information on defect dynamics and criticality by means of Kibble-Zurek formalism~\cite{Sim:2022,Sim:2023}.
%%%%%%%%%%%%%%%%%%%%%%%%%%%%%%%%%%%%%%%%%%%%%%%%%%%%%%%%%%%%%%%%%%%%%%%%%%%%%%%%%%%%%%%%%%%%%

%%%%%%%%%%%%%%%%%%%%%%%%%%%%
%%%%% ACKNOWLEDGEMENTS %%%%%
%%%%%%%%%%%%%%%%%%%%%%%%%%%%
%%%%%%%%%%%%%%%%%%%%%%%%%%%%%%%%%%%%%%%%%%%%%%%%%%%%%%%%%%%%%%%%%%%%%%%%%%%%%%%%%%%%%%%%%%%%%
\textit{Acknowledgements --}
We thank Emil Bergholtz for useful discussions.
This work was supported by the Swedish Research Council (2018-00313) and Knut and Alice Wallenberg Foundation (KAW) via the project Dynamic Quantum Matter (2019.0068).
Computation time at the High-Performance Computing Center Stuttgart (HLRS) and on the Euler cluster at the High-Performance Computing Center of ETH Zurich is gratefully acknowledged. 
%%%%%%%%%%%%%%%%%%%%%%%%%%%%%%%%%%%%%%%%%%%%%%%%%%%%%%%%%%%%%%%%%%%%%%%%%%%%%%%%%%%%%%%%%%%%%

%%%%%%%%%%%%%%%%%%%%%%
%%%%% APPENDICES %%%%%
%%%%%%%%%%%%%%%%%%%%%%
\appendix
%%%%%%%%%%%%%%%%%%%%%%%%%%%%%%%%%%%%%%%%%%%%%%%%%%%%%%%%%%%%%%%%%%%%%%%%%%%%%%%%%%%%%%%%%%%%%
\section{MCTDH-X}
\label{app:MCTDHX}
In this appendix, we briefly review the numerical method used to obtain the time evolution of the few-boson systems presented in the main text.
We employ the MultiConfigurational Time-Dependent Hartree method for indistinguishable particles implemented by the MCTDH-X software~\cite{Alon:2008,Lode:2016,Fasshauer:2016,Lin:2020,Lode:2020,MCTDHX}.
As MCTDH-X solves the many-body Schr\"{o}dinger equation in time, it is the ideal method to probe quench dynamics of interacting ultracold systems.
In fact, MCTDH-X has been applied to several dipolar-interacting systems before~\cite{Fischer:2015,chatterjee:2018,chatterjee:2019,Bera:2019,chatterjee:2020,Hughes:2023}.

MCTDH-X constructs the many-body wave function as a linear combination of time-dependent permanents
\begin{equation}
\left| \Psi(t) \right>= \sum_{\mathbf{n}}^{} C_{\mathbf{n}}(t)\vert \mathbf{n};t\rangle.
\label{many_body_wf}
\end{equation}
The permanents are constructed over $M$ time-dependent single-particle wavefunctions, called orbitals, as 
\begin{equation}
\vert \mathbf{n};t\rangle = \prod^M_{k=1}\left[ \frac{(\hat{b}_k^\dagger(t))^{n_k}}{\sqrt{n_k!}}\right] |0\rangle 
\label{many_body_wf_2}
\end{equation}
Here, $\mathbf{n}=(n_1,n_2,...,n_M)$ is the number of bosons in each orbital.
This is constraint by $\sum_{k=1}^M n_k=N$, with $N$ the total number of bosons.
Allocating $N$ bosons over $M$ orbitals, the number of permanents becomes $ \left(\begin{array}{c} N+M-1 \\ N \end{array}\right)$. 
Additionally, $|0\rangle$ is the vacuum state and $\hat{b}_k^\dagger(t)$ denotes the time-dependent operator that creates one boson in the $k$-th working orbital $\psi_k(x)$, \textit{i.e.}:
%%%
\begin{eqnarray}
	\hat{b}_k^\dagger(t)&=&\int \mathrm{d}x \: \psi^*_k(x;t)\hat{\Psi}^\dagger(x;t) \:  \\
	\hat{\Psi}^\dagger(x;t)&=&\sum_{k=1}^M \hat{b}^\dagger_k(t)\psi_k(x;t). \label{eq:def_psi}
\end{eqnarray}
%%%
The accuracy of the algorithm depends on the number of orbitals used.
For $M=1$ (a single orbital), MCTDH-X coincides with a mean-field Gross-Pitaevskii description.
For $M \rightarrow \infty$, the wave function becomes exact as the set $ \vert n_1,n_2, \dots ,n_M \rangle$ spans the complete $N$-partcle Hilbert space. 
For practical calculations, we restrict the number of orbitals to a value that is small enough to achieve convergence in the relevant observables.

Both the expansion coefficients $C_\mathbf{n}(t)$ and the working orbitals $\psi_i(x;t)$ that constitute the permanents are optimized variationally at every time step~\cite{TDVM81} to either relax the system to its ground state (imaginary time propagation), or to calculate the full dynamics of the many-body state (real time propagation).
The variational procedure occurs at the level of the many-body action obtained from the Hamiltonian of the system written in second quantization as
\begin{align} 
\hat{\mathcal{H}}&=\int dx \hat{\Psi}^\dagger(x) \left\{\frac{p^2}{2m}+V(x)\right\}\hat{\Psi}(x) \nonumber\\
&+\frac{1}{2}\int dx \hat{\Psi}^\dagger(x)\hat{\Psi}^\dagger(x')W(x,x')\hat{\Psi}(x)\hat{\Psi}(x'),
\end{align}
where $V(x)$ denotes a one-body potential and $W(x,x')$ describes two-body interactions.
For the present work, $V(x)$ is the optical lattice, while $W(x,x')$ is the sum of contact repulsion and dipole-dipole interactions.
We require the stationarity of the action with respect to variations of the time-dependent coefficients and orbitals. 
This results in a coupled set of equations of motion containing those quantities, which are then solved simultaneously.
We remark that the one particle function $ \phi_i (x,t)$ and the coefficient $C_{\bar{n}}(t)$ are variationally optimal with respect to all parameters of the many-body Hamiltonian at any time~\cite{TDVM81,variational1,variational3,variational4}. 

From the working orbitals, it is possible to calculate $N$-body reduced density matrices and obtain information about correlation in the system.
For example, the one-body reduced density matrix can be computed as
%%%
\begin{eqnarray}
\rho^{(1)}(x,x') = \sum_{kq=1}^M \rho_{kq}\psi_k(x)\psi_q(x' ),
\label{eq:red-dens-mat}
\end{eqnarray}
%%%
where
%%%
\begin{eqnarray}
\rho_{kq} = \begin{cases}
\sum_\mathbf{n} |C_\mathbf{n}|^2 n_k, \quad & k=q \\
\sum_\mathbf{n} C_\mathbf{n}^* C_{\mathbf{n}^k_q} \sqrt{n_k(n_q+1)}, \quad & k\neq q \\
\end{cases}
\end{eqnarray}
%%%
and the sum runs over all possible configurations of $\mathbf{n}$.
The quantity $\mathbf{n}^k_q$ denotes the configuration where one boson is extracted from orbital $q$ and then added to orbital $k$.
The one-particle density can be then easily computed from the one-body reduced density matrix as its diagonal elements:
%%%
\begin{equation}
	\rho(x) = \rho^{(1)}(x,x)/N
\end{equation}
%%%

Since we have access to the full many-body wave function at any time, it is a straightforward task to compute overlaps and fidelities of the type $\left<\Psi(t')\middle| \Psi(t) \right>$.
In this work, we are in particular interested in evaluating the (absolute square of the) autocorrelation function with the initial many-body state,
\begin{equation}
    \mathcal{F}(t) \equiv |\left<\Psi(0)\middle| \Psi(t) \right>|^2,
\end{equation}
which captures the stability of the initial state in time after the quantum quench, i.e. how much the time evolved state overlaps with its initial state.
%%%%%%%%%%%%%%%%%%%%%%%%%%%%%%%%%%%%%%%%%%%%%%%%%%%%%%%%%%%%%%%%%%%%%%%%%%%%%%%%%%%%%%%%%%%%%

%%%%%%%%%%%%%%%%%%%%%%%%%%%%%%%%%%%%%%%%%%%%%%%%%%%%%%%%%%%%%%%%%%%%%%%%%%%%%%%%%%%%%%%%%%%%%
\section{System parameters}

In this appendix we discuss the parameters for the simulations presented in the main text.
The system consists of $N=5$ or $N=6$ bosons in an optical lattice with lattice depth $V_0$ and wave vector $k_0 = \pi/\lambda$ parametrized as
\begin{equation}
V(x) = V_0 \sin^2 (k_0 x).
\end{equation}
We choose the wavelengths to be compatible with real experimental realizations in ultracold atomic labs, i.e. $\lambda_0 \approx 532.2$ nm.
This gives a wave vector $k_0 \approx 5.903 \times 10^6$ m$^{-1}$.
Depending on the setup, we set hard-wall barriers to restrict the optical lattice to the center-most five minima (single-filling simulations) or three minima (double-filling simulations).

\subsection{Lengths}
In MCTDH-X simulations, we choose to set the unit of length $\bar{L} \equiv \lambda_0/2 = 266.1$ nm, which makes the minima of the optical lattice (the sites in the lattice picture) appear at integer values in dimensionless units, while the maxima are located at half-integer values.
In particular, $x=0$ acts as the center of the lattice which can host an odd number of sites $S$.
In our simulations, we consider mainly two cases: an integer filling of $N=5$ bosons in $S=5$ sites, and a double filling of $N=6$ bosons in $S=3$ sites.
In both cases, we run simulations with 512 gridpoints in an interval $x \in [-4 \bar{L}, 4 \bar{L}] \approx [-1.064 \mu\mathrm{m}, 1.064 \mu\mathrm{m}]$, giving a resolution of around 4.158 nm.
We employ hard-wall boundary conditions at $x=\pm 2.5$ ($S=5$) and $\pm 1.5$ ($S=3$) to clearly demarcate the end of the optical lattice.

\subsection{Energies}
The unit of energy $\bar{E}$ is effectively defined in terms of the recoil energy of the optical lattice, i.e. $E_r \equiv \frac{\hbar^2 k_0^2}{2m} \approx 2.204 \times 10^{-30}$ J with $m \approx 51.941$ Da the mass of $^{52}$Cr atoms, a dipolar magnetic species amply used in experiments~\cite{Griesmaier:2005, Lahaye:2008}.
More specifically, we define the unit of energy as 
$\bar{E} \equiv \frac{\hbar^2}{m \bar{L}^2} = \frac{2 E_r}{\pi^2} \approx 4.466 \times 10^{-31}$ J.

In typical experiments, the optical lattice depth is varied in regimes of up to several tens of recoil energies (in particular when lattice physics is quantum simulated).
In our simulations, we probe similar regimes of $V_0 = 100 \bar{E} \approx 20 E_r$. 
The on-site interactions are kept at fixed magnitude $|g_0| = 10 \bar{E} \approx 2 E_r$, unless otherwise stated.
During the quench, we change only their sign (from repulsive to attractive).
When we turn on dipolar interactions, we probe regimes up to strength $|g_d| = 2.0 \bar{E} \approx 0.4 E_r$, since the physics does not change qualitatively with stronger dipolar interactions.
Empirically, we find that around $|g_d| \approx 0.2 E_r$ -- i.e. around 10\% of the contact interaction strength -- is sufficient to condense all bosons into a central droplet despite the strong initial repulsions.

\subsection{Time}
The unit of time is also defined from the unit of length, as 
$\bar{t} \equiv \frac{m \hat{L}^2}{\hbar} = \frac{m \lambda_0^2}{\hbar} = 0.5903 \times 10^{-4}$ s $=59.03$ $\mu$s.
In our simulations, we ran all time evolutions up until at least $t \approx 5-6 \bar{t} \approx 300-350$ $\mu$s.
However, for the fidelity calculations, we typically let the time evolution run for longer than $t \approx 20 \bar{t} \approx 1.2$ ms.

%%%%%%%%%%%%
\begin{table}
\centering
\begin{tabular}{ || c | c || }
\hline \hline
Quantity & MCTDH-X units  \\
\hline \hline
unit of length &  $\bar{L} = \lambda_0/2 = 266.1$ nm \\
\hline
opt. latt. sites & at -2, -1, 0, 1, 2 \\
\hline
unit of energy & $\bar{E} = \frac{\hbar^2}{m \bar{L}^2} =\frac{2E_r}{\pi^2} \approx 4.466 \times 10^{-31}$ J \\
\hline
potential depth & $V=100.0 \bar{E} \approx 20 \: E_r$ \\
\hline
on-site repulsion & $|g_0| = 10.0 \bar{E} \approx 2 \: E_r$ \\
\hline
dipolar interaction & e.g. $g_d=-1.0 \bar{E} \approx -0.2 \: E_r$  \\
\hline
unit of time & $\bar{t} \equiv \frac{m\bar{L}^2}{\hbar} = 59.03$ $\mu$s \\
\hline \hline
\end{tabular}
\caption{Units used in MCTDH-X simulations. $E_r=\frac{\hbar^2 k_0^2}{2m}$ is the recoil energy.}
\end{table}
%%%%%%%%%%%
%%%%%%%%%%%%%%%%%%%%%%%%%%%%%%%%%%%%%%%%%%%%%%%%%%%%%%%%%%%%%%%%%%%%%%%%%%%%%%%%%%%%%%%%%%%%%

%%%%%%%%%%%%%%%%%%%%%%%%%%%%%%%%%%%%%%%%%%%%%%%%%%%%%%%%%%%%%%%%%%%%%%%%%%%%%%%%%%%%%%%%%%%%%
\section{Orbital convergence}
\label{app:orbital-conv}

In this section we discuss how our results depend on the number of orbitals $M$ employed in the computations.
The accuracy of the MCTDH-B ansatz crucially depends on the choice of $M$.
Strictly speaking, the wave function becomes exact only in the limit $M \rightarrow \infty$.
However, very often it is possible to achieve converged, \emph{numerically} exact results while employing a finite number of orbitals.
An important criterion used to quantify the correctness of the calculated observables is orbital convergence, which requires the main features of the system at hand to stabilize as the number of orbitals used in the computations is progressively increased.
To that end, we have probed convergence both in the initial state preparation and in the quench dynamics by repeating the main text calculations with $M=12$ to $M=15$ orbitals.

Let us preface the discussion of the convergence results by saying that the many-body problem we are currently investigating is extraordinarily challenging for several reasons:
(1) sudden quench procedures are incredibly disruptive to the wave function structure and to the computation of time evolutions, because of the deep orbital reorganization that has to happen in real time.
(2) This is especially true when we consider very strong interactions (several recoil energies) as in our case.
(3) Attractive bosons are particularly difficult to converge with the MCTDH-B method. 
The issue is more severe with short-range interaction in free space~\cite{Cosme:2016}, but can be mitigated for finite-range interaction~\cite{Alon:2008}.
(4) there is a computational tradeoff between employing a fine spatial resolution and enabling a faster integration of the many-body Schr\"{o}dinger equation to reach longer times. 
In this work, we have chosen to focus on probing the longest times possible with the highest accuracy. 
This requires us to employ a coarser spatial resolution, which may sometimes result in slight asymmetries and kinks in the density profile.

\begin{figure}
\centering
\includegraphics[width=\columnwidth]{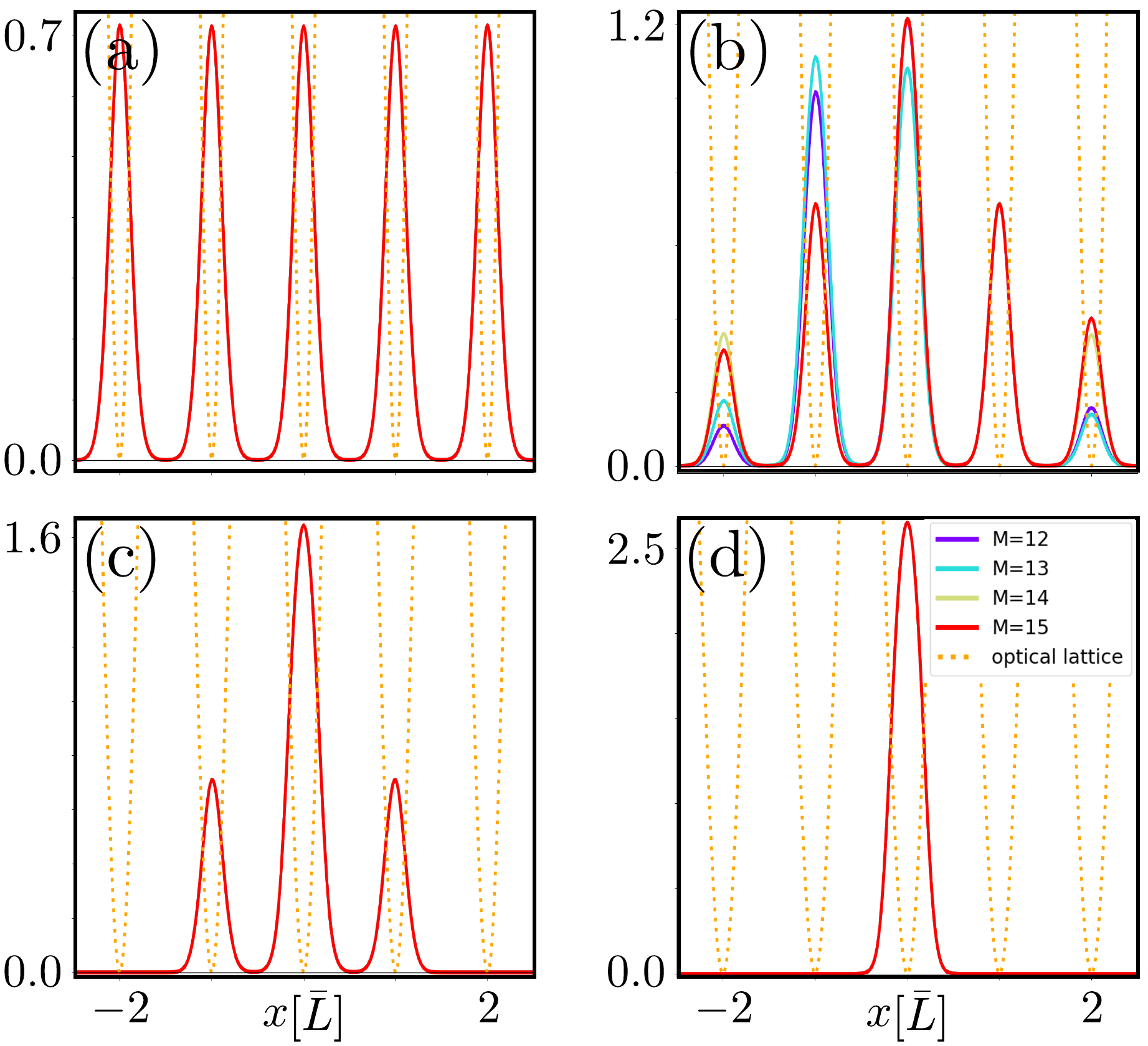}
\caption{
Orbital convergence of the initial states (solid line) and optical lattice (dashed line, scaled down by 20$\times$) used in the quench protocol for $N=5$ bosons in $S=5$ sites (unit filling) with 
(a) $g_d=-0.1 E_r$, 
(b) $g_d=-0.18 E_r$,
(c) $g_d=-0.19 E_r$, 
(d) $g_d=-0.4 E_r$.
The other parameters are $V_0 \approx 20 E_r$ and $|g_0| \approx 2 E_r$.
}
\label{fig:initial-states-single-occ-M-comparison}
\end{figure}

\begin{figure}
\centering
\includegraphics[width=\columnwidth]{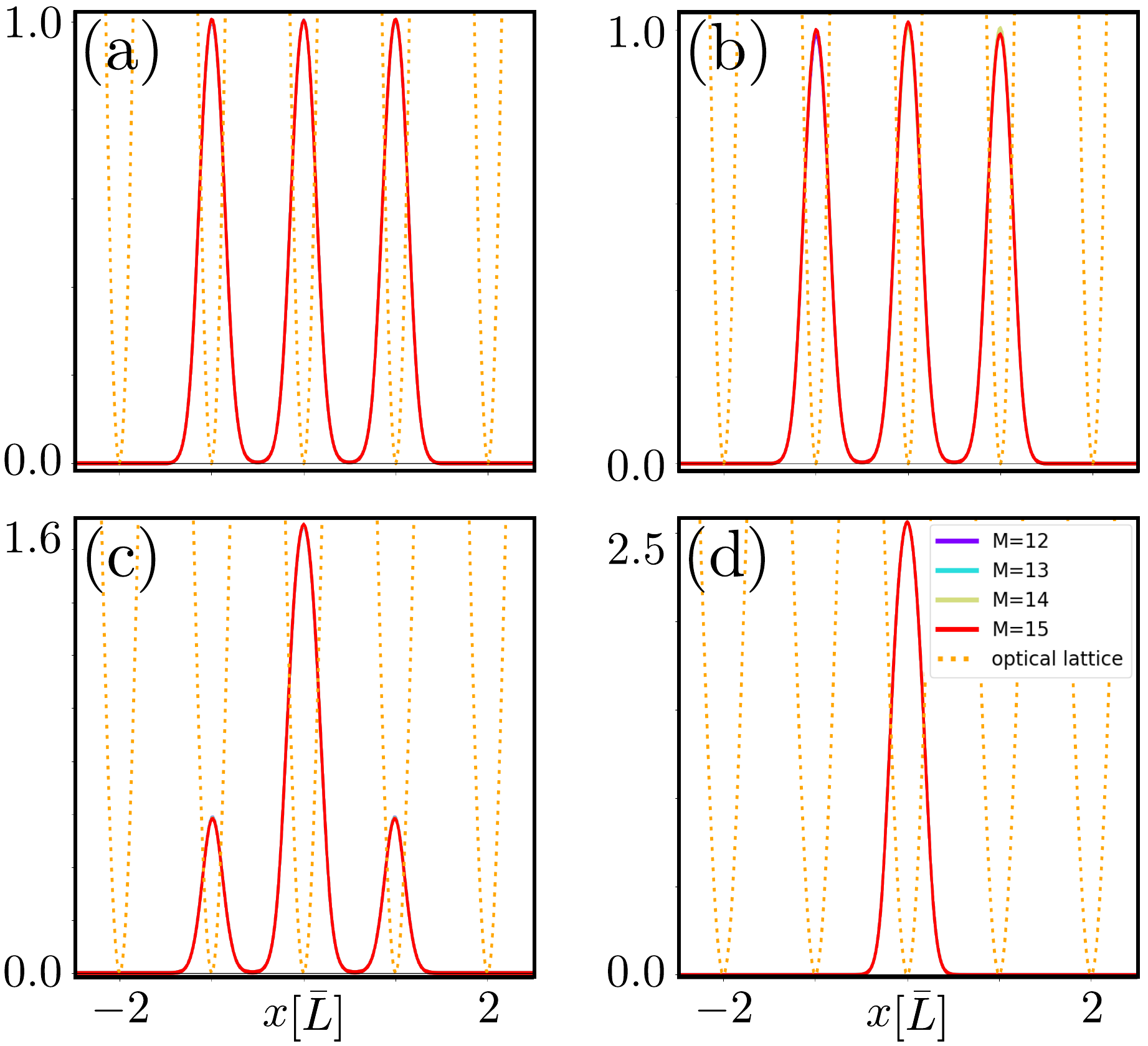}
\caption{
Orbital convergence of the initial states (solid line) and optical lattice (dashed line, scaled down by 20$\times$) used in the quench protocol for $N=6$ bosons in $S=3$ sites (double filling) with 
(a) $g_d=-0.16 E_r$, 
(b) $g_d=-0.17 E_r$, 
(c) $g_d=-0.19 E_r$,
(d) $g_d=-0.4 E_r$.
The other parameters are $V_0 \approx 20 E_r$ and $|g_0| \approx 2 E_r$.
}
\label{fig:initial-states-double-occ-M-comparison}
\end{figure}

We first present the density profile for the initial states used in the quench protocol and studied in the main text.
The initial states for the unit filling case obtained with different numbers of orbitals can be seen in Fig.~\ref{fig:initial-states-single-occ-M-comparison}, while the ones for the double filling case are shown in Fig.~\ref{fig:initial-states-double-occ-M-comparison}.
We remark that close to the transition between the Mott and cluster phases ($g_d \approx 0.2 E_r$) there is a strong competition between lattice localization, contact repulsion, and dipolar attraction.
As we can see, around the phase transition there are some fluctuations around the symmetric structure for lower values of $M$, see Fig.~\ref{fig:initial-states-single-occ-M-comparison}(b).
However, employing $M=15$ yields a safely converged initial state for all the parameter ranges probed in this work.

%%%%%%%%%%%%%%%%%%%%%%%%%%%%%%%%%%%%%%%%%%%%
%%%%%%%%%%%%%%%%%%%%%%%%%%%%%%%%%%%%%%%%%%%%

%%%%%%%%%%%%%%%%%%%%%%%%%%%%%%%%%%%%%%%%%%%%
%%%%%%%%%%%%%%%%%%%%%%%%%%%%%%%%%%%%%%%%%%%%

We now discuss how the dynamics is affected by adopting an increasing number of orbitals in the MCTDH decomposition.
For brevity, we only consider the single-filling case with $g_d=-0.18 E_r$, as it appears to be the harder case to converge even for the initial state.
We nevertheless verified that the dynamics is converged in the number of orbitals for other values of $g_d$ and double filling.

\begin{figure}
\centering
\includegraphics[width=\columnwidth]{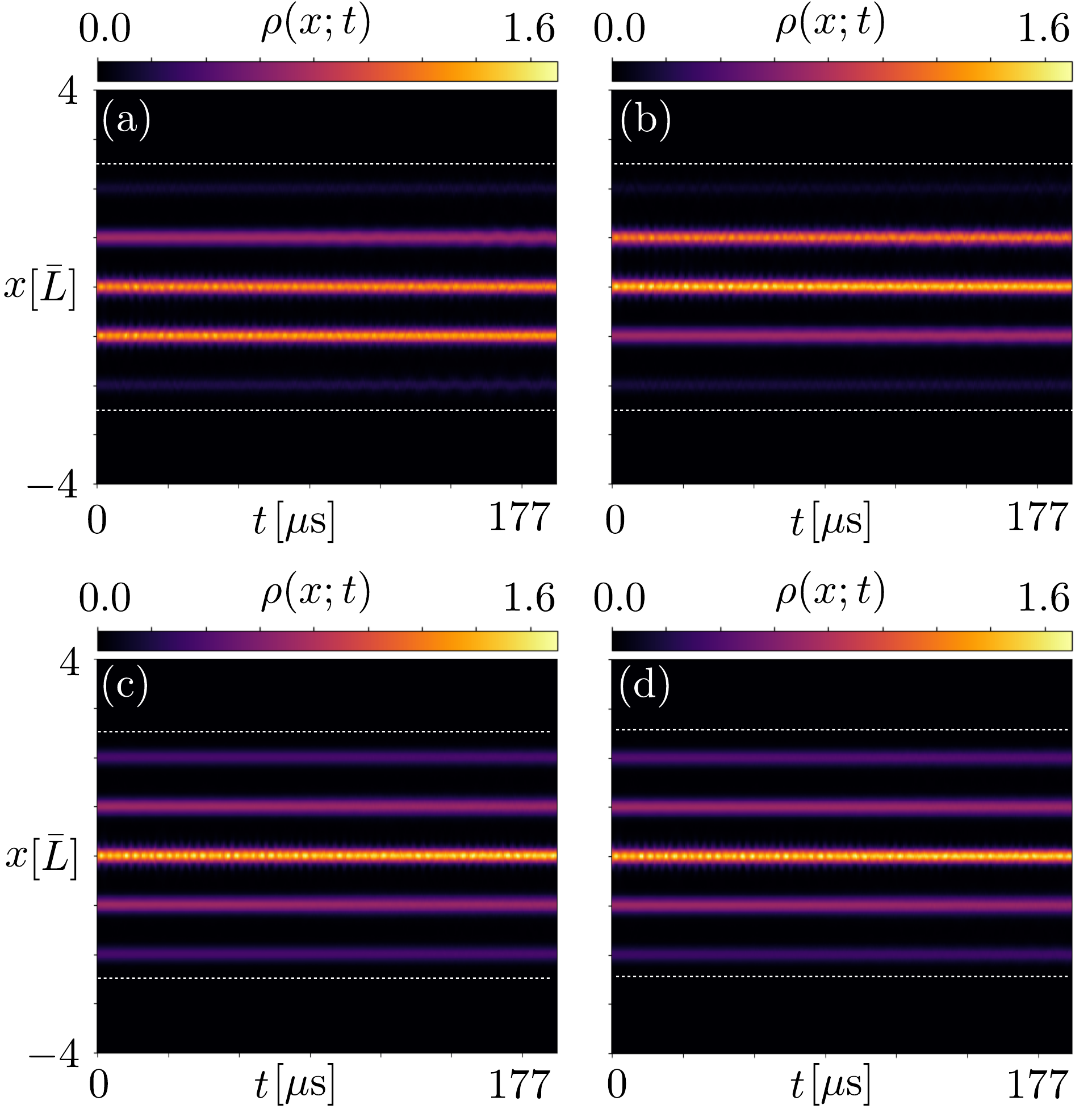}
\caption{
Post quench density dynamics for $N=5$ bosons in $S=5$ sites (single filling) with dipolar interaction strength $g_d=-0.18 E_r$, plotted for increasing orbital number $M$:
(a) $M=12$,
(b) $M=13$,
(c) $M=14$,
(d) $M=15$.
The other parameters are $V_0 \approx 20 E_r$ and $|g_0| \approx 2 E_r$.
}
\label{fig:single-occ-M-comparison}
\end{figure}

The density dynamics after the quench is illustrated in Fig.~\ref{fig:single-occ-M-comparison} for an increasing number of orbitals up to $M=15$ (corresponding to 3 orbitals per site on average).
While we do observe some slight asymmetries with a lower number of orbitals, the results for $M=14$ and $M=15$ very closely resemble each other, indicating that the density dynamics is converged at $M=15$.

\begin{figure}
\centering
\includegraphics[width=\columnwidth]{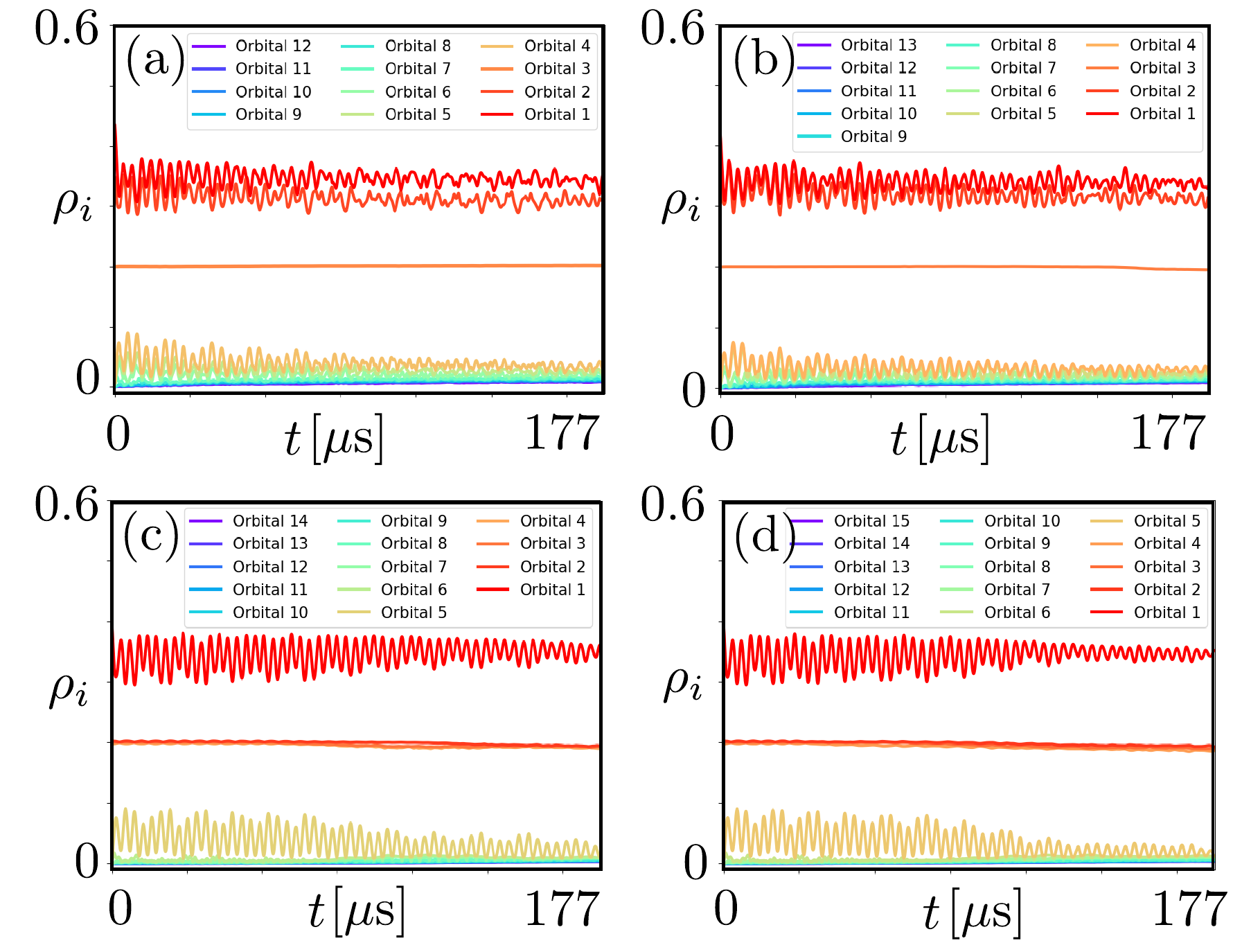}
\caption{
Dynamics of orbital occupation for $N=5$ bosons in $S=5$ sites (single filling) with dipolar interaction strength $g_d=-0.18 E_r$, plotted for increasing orbital number $M$:
(a) $M=12$,
(b) $M=13$,
(c) $M=14$,
(d) $M=15$.
The other parameters are $V_0 \approx 20 E_r$ and $|g_0| \approx 2 E_r$.
}
\label{fig:single-occ-orbital-occ-M-comparison}
\end{figure}

Concomitantly, we observe that the orbital occupation saturates with increasing number of orbitals, as evinced from Fig.~\ref{fig:single-occ-orbital-occ-M-comparison}.
In fact, only up to eight orbitals appear to have a macroscopic population.
We have verified that the additional orbitals (9-15) systematically contribute to less than 1\% to the total many-body wave function throughout the entire time evolution.
Since adding more orbitals not only does not change the overall dynamics, but also results in negligible occupation of the newly introduced orbitals, we can declare the overall dynamics converged.

%%%%%%%%%%%%%%%%%%%%%%%%%%%%%%%%%%%%%%%%%%%%
%%%%%%%%%%%%%%%%%%%%%%%%%%%%%%%%%%%%%%%%%%%%

%%%%%%%%%%%%%%%%%%%%%%%%%%%%%%%%%%%%%%%%%%%%
%%%%%%%%%%%%%%%%%%%%%%%%%%%%%%%%%%%%%%%%%%%%
%%%%%%%%%%%%%%%%%%%%%%%%%%%%%%%%%%%%%%%%%%%%%%%%%%%%%%%%%%%%%%%%%%%%%%%%%%%%%%%%%%%%%%%%%%%%%

%%%%%%%%%%%%%%%%%%%%%%%%%%%%%%%%%%%%%%%%%%%%%%%%%%%%%%%%%%%%%%%%%%%%%%%%%%%%%%%%%%%%%%%%%%%%%
\section{Autocorrelation for $M=15$}
\label{app:autocorr}

\begin{figure}[t]
\centering
\includegraphics[width=\columnwidth]{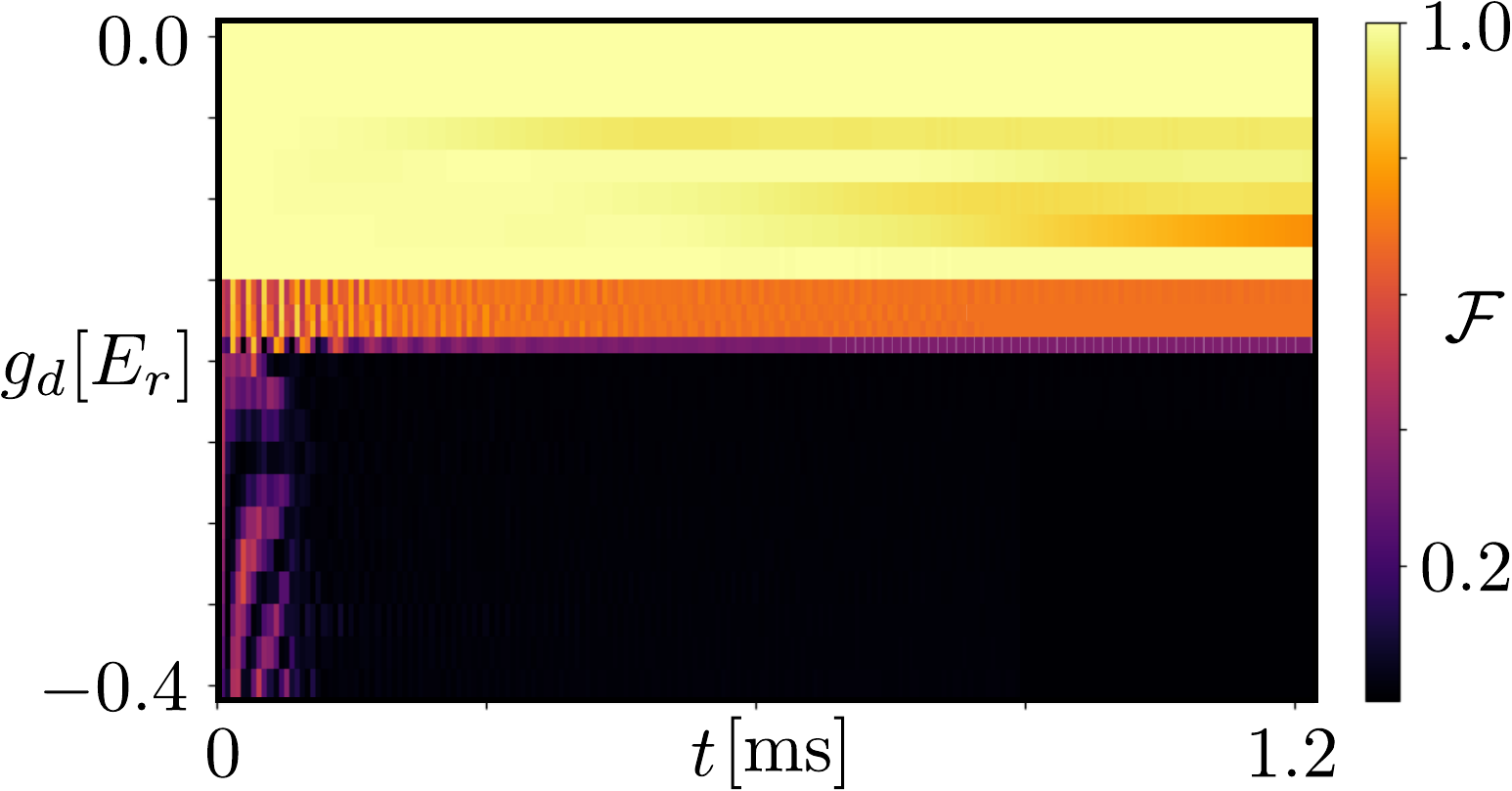}
\caption{
Dynamics of the autocorrelation function $\mathcal{F}(t)$ post quench for $N=5$ bosons and $M=15$ orbitals in $S=5$ sites (unit filling), plotted in the entire range of probed dipolar attractions $g_d$ from $0.0$ to $-0.4 E_r$.
The other parameters are $V_0 \approx 20 E_r$ and $|g_0| \approx 2 E_r$.
}
\label{fig:fidelity-M-15}
\end{figure}

In this appendix, we show results for the calculation of the autocorrelation function with $M=15$ orbitals.
The autocorrelation function is mapped as a function of the interaction strength $g_d$ and time $t$ in Fig.~\ref{fig:fidelity-M-15}.
The behavior of the autocorrelation function is essentially equivalent to the one obtained for $M=12$ orbitals and presented in the main text.
This ensures that our results with $M=12$ orbitals are already reliable.

The autocorrelation function separates the dipolar interaction strengths into three regimes.
A first regime persists down to $g_d \approx -0.1 E_r$ and is characterized by exceptional long-time stability of the quenched state.
A second regime can be observed between $g_d \approx -0.1 E_r$ and $g_d \approx -0.19 E_r$ and exhibits a reduction of stability with values of autocorrelation dropping all the way down to 0.2 for the stronger interactions.
A third regime occurs below $g_d \approx -0.19 E_r$ and reveals a complete collapse of the initial state as evinced by autocorrelation values close to 0.

%%%%%%%%%%%%%%%%%%%%%%%%%%%%%%%%%%%%%%%%%%%%
% %%%%%%%%%%%%%%%%%%%%%%%%%%%%%%%%%%%%%%%%%%%%

%%%%%%%%%%%%%%%%%%%%%%%%%%%%%%%%%%%%%%%%%%%%%%%%%%%%%%%%%%%%%%%%%%%%%%%%%%%%%%%%%%%%%%%%%%%%%

%%%%%%%%%%%%%%%%%%%%%%%%%%%%%%%%%%%%%%%%%%%%%%%%%%%%%%%%%%%%%%%%%%%%%%%%%%%%%%%%%%%%%%%%%%%%%
\section{Other measures of correlation}
\label{app:obs}
\begin{figure}[t]
\centering
\includegraphics[width=\columnwidth]{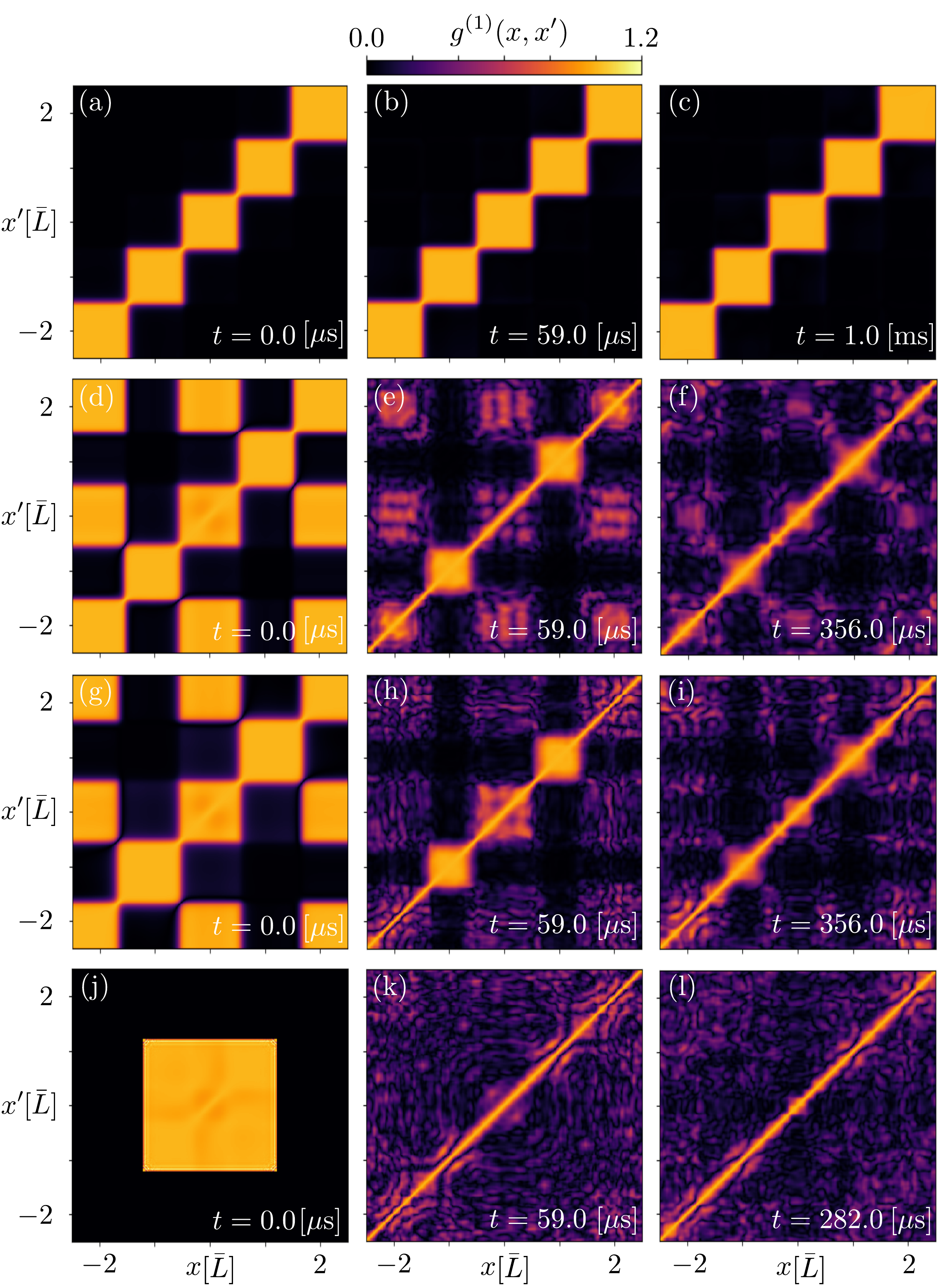}
\caption{
Dynamics of the Glauber one-body correlation function $g^{(1)}(x,x')$ post quench for $N=5$ bosons and $M=15$ orbitals with 
(a)-(c) $g_d=-0.16 E_r$, 
(d)-(f) $g_d=-0.18 E_r$,
(g)-(i) $g_d=-0.19 E_r$, 
(j)-(l) $g_d=-0.4 E_r$.
The other parameters are $V_0 \approx 20 E_r$ and $|g_0| \approx 2 E_r$.
}
\label{fig:g1-single}
\end{figure}

\begin{figure}[t]
\centering
\includegraphics[width=\columnwidth]{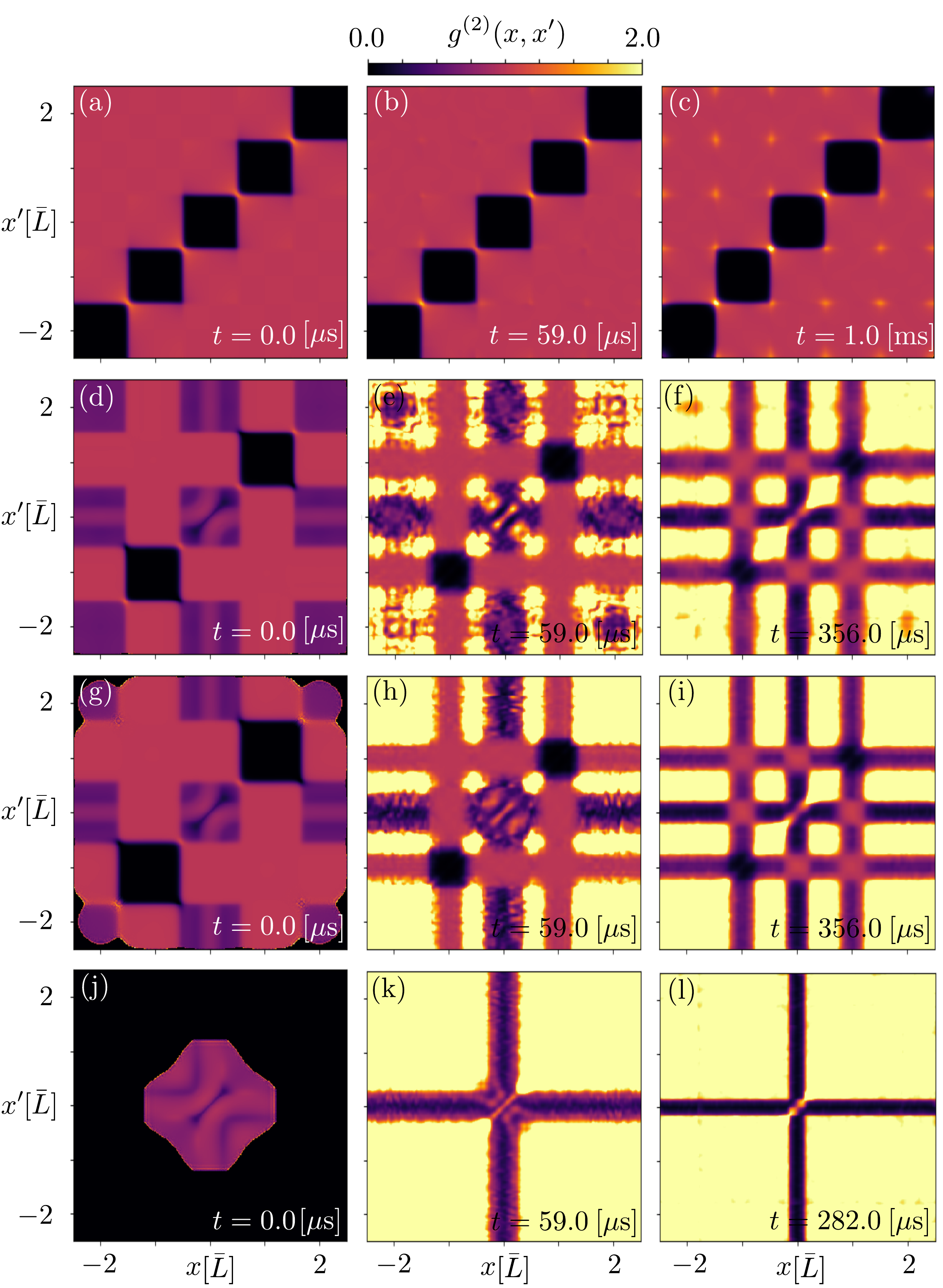}
\caption{
Dynamics of the Glauber two-body correlation function $g^{(2)}(x,x')$ post quench for $N=5$ bosons and $M=15$ orbitals with 
(a)-(c) $g_d=-0.16 E_r$, 
(d)-(f) $g_d=-0.18 E_r$,
(g)-(i) $g_d=-0.19 E_r$, 
(j)-(l) $g_d=-0.4 E_r$.
The other parameters are $V_0 \approx 20 E_r$ and $|g_0| \approx 2 E_r$.
}
\label{fig:g2-single}
\end{figure}

\begin{figure}
\centering
\includegraphics[width=\columnwidth]{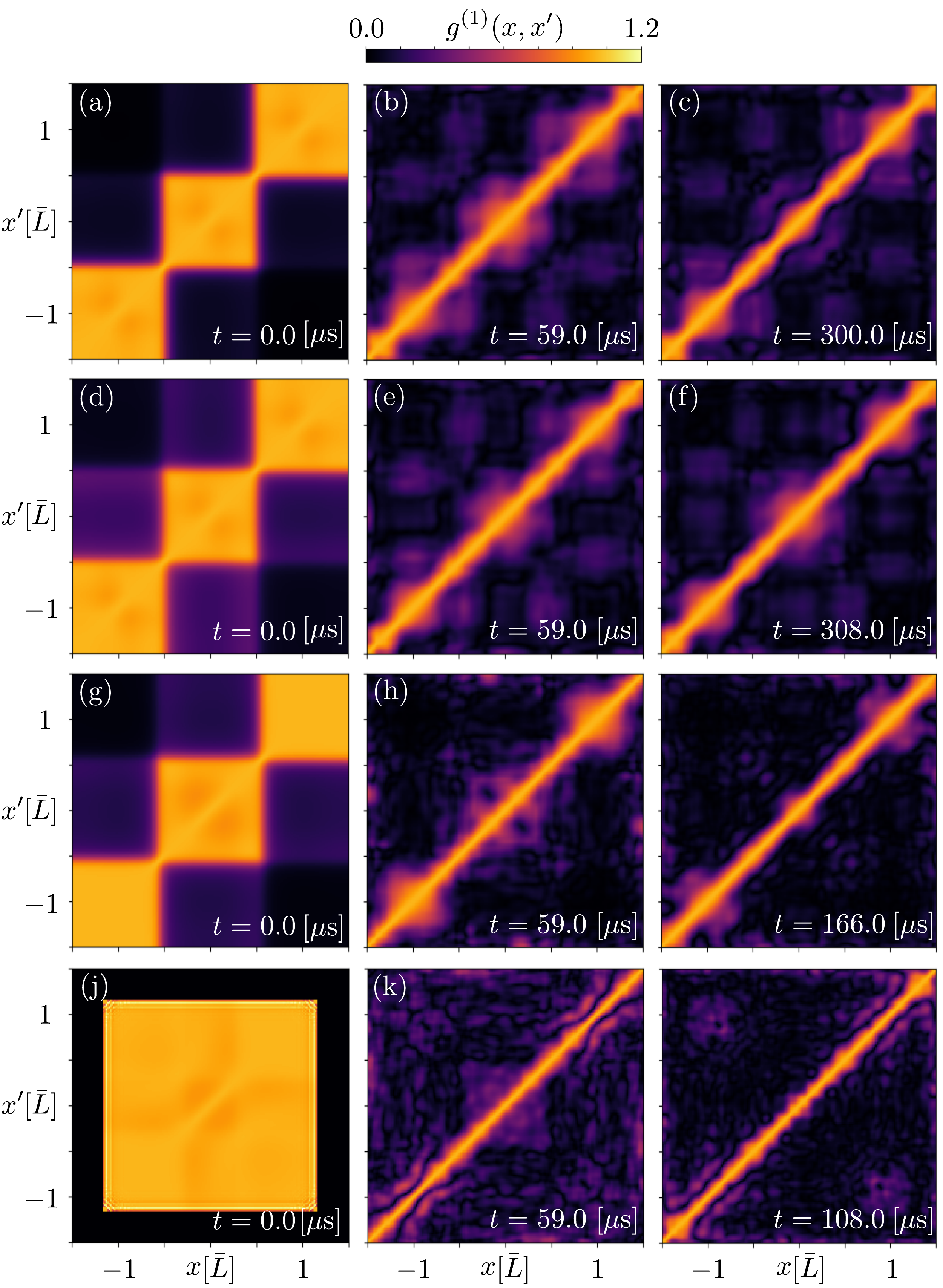}
\caption{
Dynamics of the Glauber one-body correlation function $g^{(1)}(x,x')$ post quench for $N=6$ bosons and $M=15$ orbitals in $S=3$ sites (double filling) with 
(a)-(c) $g_d=-0.16 E_r$, 
(d)-(f) $g_d=-0.17 E_r$, 
(g)-(i) $g_d=-0.19 E_r$,
(j)-(l)  $g_d=-0.4 E_r$.
The other parameters are $V_0 \approx 20 E_r$ and $|g_0| \approx 2 E_r$.
}
\label{fig:g1-double}
\end{figure}

\begin{figure}
\centering
\includegraphics[width=\columnwidth]{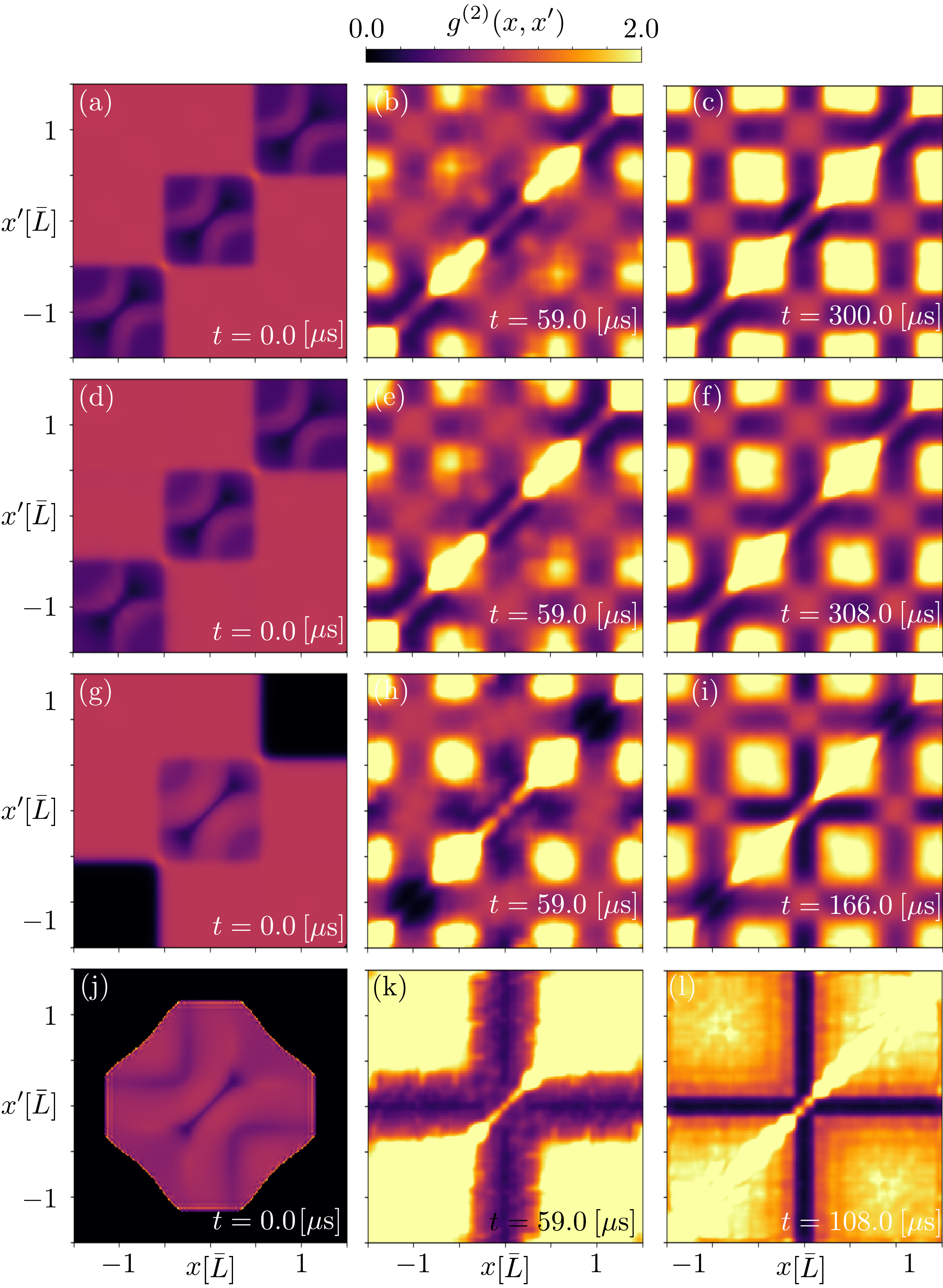}
\caption{
Dynamics of the Glauber two-body correlation function $g^{(2)}(x,x')$ post quench for $N=6$ bosons and $M=15$ orbitals in $S=3$ sites (double filling) with 
(a)-(c) $g_d=-0.16 E_r$, 
(d)-(f) $g_d=-0.17 E_r$, 
(g)-(i) $g_d=-0.19 E_r$,
(j)-(l)  $g_d=-0.4 E_r$.
The other parameters are $V_0 \approx 20 E_r$ and $|g_0| \approx 2 E_r$.
}
\label{fig:g2-double}
\end{figure}

In this appendix, we present the results for the one-body and two-body Glauber correlation functions $g^{(1)}(x,x') = \frac{\rho^{(1)}(x,x')}{N \sqrt{\rho(x) \rho(x')}}$ and $g^{(2)}(x,x')= \frac{\rho^{(2)}(x,x')}{N^2 \rho(x) \rho(x')}$. 
These quantities, shown in Figs.~\ref{fig:g1-single} to \ref{fig:g2-double}, provide a more comprehensive picture of the correlations in the system by normalizing the one-body density $\rho^{(1)}(x,x')$ and two-body density $\rho^{(2)}(x,x')$ with the particle number and density, although they conclude with the same physics drawn in the main text. 
We find that the dynamics of first-order and second-order coherence is strongly related to the dynamical instability of the density.
The parameters for post quench dynamics remain the same as described in the main text. 

The results for the unit filling case are presented in Figs.~\ref{fig:g1-single} and \ref{fig:g2-single}. 
For the prequench states with interactions up to $g_d=-0.16E_r$ [Fig.~\ref{fig:g1-single}(a)], the diagonal of the first order correlation function shows five completely separated coherent regions (bright lobes) where $|g^{(1)}|^{2}$ $\approx 1$. 
The coherence of bosons within the same well is maintained whereas the off-diagonal correlation ($x \neq x^{\prime}$) vanishes. 
This describes the five-fold fragmented, fully localized Mott phase. 
The corresponding $g^{(2)}$ in Figs.~\ref{fig:g2-single}(a) shows five completely extinguished lobes (correlation holes), i.e. the probability of finding two bosons in the same place ($x=x^{\prime})$ is zero. 
On the other hand, second order coherence is maintained between the wells. 
After the quench, the many body state retains the initial correlation profile for very long times (we performed simulations until 1 ms), as seen in Figs.~\ref{fig:g1-single}(b)-(c) and \ref{fig:g2-single}(b)-(c). 
All the other initial state at stronger interactions ($g_d = -0.18 E_r$, $g_d=-0.19 E_r$, and $g_d=-0.4 E_r$) are eventually destabilized by the sudden quench.
This behavior is exemplified by the disappearance of diagonal correlation in one-body [Fig.~\ref{fig:g1-single}(d)-(l)] and disruption of the correlation hole in two-body [Fig.~\ref{fig:g2-single} (d)-(l)] coherence. 
This behavior is analogous to what already observed in the reduced one-body and two-body density matrices in the main text.

In Figs.~\ref{fig:g1-double} and \ref{fig:g2-double}, we plot the one- and two-body correlation functions $g^{(1)}$ and $g^{(2)}$ for the double filling case where the Mott phase is perturbed by weak attractive long-range interaction. 
As already seen for the reduced one-body and two-body density matrices in the main text, every initial state -- even with very weak interactions -- eventually collapses after the sudden quench.
The collapse of correlations occurs quite rapidly.
Already at $t = 59 \mu$s we observe a dramatic disappearance of correlation patterns, both at the one-body and at the two-body level.

%%%%%%%%%%%%%%%%%%%%%%%%%%%%%%%%%%%%%%%%%%%%
% %%%%%%%%%%%%%%%%%%%%%%%%%%%%%%%%%%%%%%%%%%%%

%%%%%%%%%%%%%%%%%%%%%%%%%%%%%%%%%%%%%%%%%%%%
%%%%%%%%%%%%%%%%%%%%%%%%%%%%%%%%%%%%%%%%%%%%

%%%%%%%%%%%%%%%%%%%%%%%%%%%%%%%%%%%%%%%%%%%%
%%%%%%%%%%%%%%%%%%%%%%%%%%%%%%%%%%%%%%%%%%%%

% %%%%%%%%%%%%%%%%%%%%%%%%%%%%%%%%%%%%%%%%%%%%
% %%%%%%%%%%%%%%%%%%%%%%%%%%%%%%%%%%%%%%%%%%%%

% %%%%%%%%%%%%%%%%%%%%%%%%%%%%%%%%%%%%%%%%%%%%%%%%%%%%%%%%%%%%%%%%%%%%%%%%%%%%%%%%%%%%%%%%%%%%%

%%%%%%%%%%%%%%%%%%%%%%%%%%%%%%%%%%%%%%%%%%%%%%%%%%%%%%%%%%%%%%%%%%%%%%%%%%%%%%%%%%%%%%%%%%%%%

%%%%%%%%%%%%%%%%%%%%%%%%%%%%%%%%%%%%%%%%%%%%%%%%%%%%%%%%%%%%%%%%%%%%%%%%%%%%%%%%%%%%%%%%%%%%%
\bibliography{biblio}
%%%%%%%%%%%%%%%%%%%%%%%%%%%%%%%%%%%%%%%%%%%%%%%%%%%%%%%%%%%%%%%%%%%%%%%%%%%%%%%%%%%%%%%%%%%%%

\end{document}